\documentclass[14pt]{article}
\usepackage{epsf}
\usepackage{color}
\usepackage{framed}
\usepackage{amsmath,amsfonts,amsbsy,amssymb}
\usepackage{indentfirst}
\usepackage{mathtools}
\usepackage{arydshln} 
\usepackage{framed}

\newcommand{\nn}{\nonumber}

\def\be{\begin{equation}}
\def\ee{\end{equation}}
\def\bse{\begin{subequations}}
\def\ese{\end{subequations}}
\def\bal{\begin{align}}
\def\ealn{\end{align}}
\def\tr{\text{tr}}
\def\bs{\boldsymbol}

\begin{document}

\begin{titlepage}

\def\slash#1{{\rlap{$#1$} \thinspace/}}

\begin{flushright} 

\end{flushright} 

\vspace{0.1cm}

\begin{Large}
\begin{center}

{\bf  Spin-entangled  Squeezed State  on   a Bloch Four-hyperboloid  }
\end{center}
\end{Large}

\vspace{1cm}

\begin{center}
{\bf Kazuki Hasebe}   \\ 
\vspace{0.5cm} 
\it{
National Institute of Technology, Sendai College,  
Ayashi, Sendai, 989-3128, Japan} \\ 

\vspace{0.5cm} 

{\sf
khasebe@sendai-nct.ac.jp} 

\vspace{0.8cm} 

{\today} 

\end{center}

\vspace{1.0cm}

\begin{abstract}
\noindent

\baselineskip=18pt

 The Bloch hyperboloid $H^2$  
 underlies the  quantum geometry of the original $SO(2,1)$ squeezed states. 
In \cite{Hasebe-2019}, 
the author utilized a non-compact 2nd Hopf map  and a Bloch four-hyperboloid $H^{2,2}$   
to explore an $SO(2,3)$ extension of the squeezed states. In the present paper, we further pursue the idea to derive an $SO(4,1)$ version of  squeezed vacuum   
based on 
the other Bloch four-hyperboloid $H^4$.  We show that the obtained $SO(4,1)$ squeezed vacuum  is a particular four-mode squeezed state  not quite similar to the previous $SO(2,3)$ squeezed vacuum. In view of  the Schwinger's formulation of angular momentum,  
%
 the $SO(4,1)$ squeezed vacuum is interpreted as a superposition  of an infinite number of  
 maximally entangled spin-pairs of all integer spins.  
We clarify  basic properties of the $SO(4,1)$ squeezed vacuum, such as  von Neumann entropy of spin entanglement, spin correlations  and uncertainty relations with emphasis on  their distinctions  to  the original $SO(2,1)$ case.

\end{abstract}

\end{titlepage}

\newpage 

\tableofcontents

\newpage 

\section{Introduction}

Squeezed state \cite{Yuen-1976,Hollenhorst-1979, Caves-1981, Walls-1983, Schumaker-Caves-1985-1,Schumaker-Caves-1985-2} is a particularly important quantum state in  quantum optical information  and   practical applications (see \cite{Drummond-Ficek-2003, Bachor-Ralph-2019, Schnabel-2017} as nice reviews.)   
In quantum optical information,   two-mode squeezed light realizes  an entangled  state of photons and it has been employed  in state-of-the-art experiments  of quantum mechanics (see \cite{Gerry-Knight-2004} for instance and references therein).   Also  in practical application to quantum telecommunication,  
the squeezed state  has attracted a lot of attention due to its intrinsic anisotropic property of quantum noise.  
The  uncertainty region of squeezed state is not isotropic unlike the coherent state (laser) and an unfavorable quantum noise   is reduced  in a specific direction.

Up to now, many theoretical extensions of the squeezed state have been proposed \cite{Milburn-1984,Bishop-Vourdas-1988,Ma-Rhodes-1990,Han-Kim-Noz-Yeh-1993,Arvind-Dutta-Mukunda-Simon-1995-1,Arvind-Dutta-Mukunda-Simon-1995-2, Yukawa-Nemoto-2016, Svozil-1990,Schmitt-Mufti-1991,Schmitt-1993}.  
In an attempt of  extending the formalism of the squeezed state,  mathematical  aesthetics and  
 logically natural extension  should be appreciated.  One  fascinating way to formulate a theory of physics may resort to geometry. 
 Needless to say,  most successful theories of physics in this approach may be  Einstein's general relativity of  Riemann geometry   and Yang-Mills theory of  
fibre-bundle geometry.      
Also in  quantum information theory,  there is a reasonable fact that we can  believe that   geometric approach plays a key role, since  qubit, the  fundamental object in quantum information,   is described by  the geometry of  Bloch sphere associated with the Hopf map \cite{Bloch-1946, Mosseri-Dandoloff-2001},  and geometric structure   seems to be  inherent in the formulation.      

As Bloch sphere provides a geometric description of  qubit, Bloch hyperboloid plays a similar role in description of  squeezed state [see \cite{Hasebe-2019} and references therein].  
Mathematically,  Bloch sphere realizes the basemanifold of the 1st Hopf map, and similarly Bloch hyperboloid denotes the basemanifold of the non-compact 1st Hopf map. 
For 2D hyperboloids, we have two kinds of hyperboloids, two-leaf hyperboloid and one-leaf hyperboloid, which are related to the two kinds of the non-compact 1st Hopf maps: 
\be
 H^{2}\simeq H^{2,1}/S^1,~~~~~ H^{1,1}\simeq H^{2,1}/H^1.   \label{noncomphopfs}
\ee
In both cases, the total bundle space is  $H^{2,1}$. The base-manifolds are $H^2$ and $H^{1,1}$   with  compact fibre $S^1$ and  non-compact fibre $H^1$ respectively.  The geometric origin of the original squeezed state is accounted for by the left of (\ref{noncomphopfs}) with the Bloch hyperboloid $H^2$.  Interestingly, the 1st non-compact Hopf maps have their higher dimensional cousins, $i.e.$ the 2nd  and  3rd non-compact Hopf maps \cite{Hasebe-2009, Hasebe-2012}. 
This observation led the author to propose an extension of squeezed state based  on  the geometry of higher dimensional hyperboloids  \cite{Hasebe-2019}.\footnote{The non-compact Hopf maps have already been applied to  several subjects, such as twistorial quantum Hall effect \cite{Hasebe-2010-2}, non-Hermitian topological insulators \cite{Esaki-Sato-Hasebe-Kohmoto-2011,Sato-Hasebe-Esaki-Kohmoto-2011}, and cosmological models of matrix model \cite{Steinacker-2017,Sperling-Steinacker-2018,Stern-Xu-2018}. 
Higher dimensional hyperboloid $H^{p,q}$ is  defined as 
\be
\sum_{i=1}^p x^ix^i -\sum_{i=p+1}^{p+q+1}x^ix^i =-1. 
\ee
As a coset space, $H^{p,q}$ is represented as 
\be
H^{p,q} ~\simeq ~SO(p,q+1)/SO(p,q),  
\ee
and the topology is 
\be
H^{p,q} ~\simeq~\mathbb{R}^p\otimes S^{q}. 
\ee
In low dimensions, $H^{p,q}$ realizes 
 Anti-de Sitter,  de Sitter and Euclidean anti-de Sitter spaces: 
\begin{align}
&H^{2,0}=EAdS^2, ~~H^{1,1}=dS^2=AdS^2, \nn\\
&H^{4,0}=EAdS^4, ~~H^{3,1}=AdS^4 , ~~H^{1,3}=dS^4.  
\end{align}
}  
As a higher dimensional analogue of  (\ref{noncomphopfs}), we have 
the 2nd non-compact Hopf maps:
\be
 H^{4}\simeq H^{4,3}/S^3,~~~~~~~ H^{2,2}\simeq H^{4,3}/H^{2,1}.   \label{noncomphopfs2nd}
\ee
As in (\ref{noncomphopfs}), the total space in both cases of (\ref{noncomphopfs2nd}) is  given by $H^{4,3}$, and   $H^4$ and $H^{2,2}$ respectively realize the base-manifolds  with  compact fibre $S^3$ and  non-compact fibre 
$H^{2,1}$. 
 We utilized  $H^{2,2}$ (the right of (\ref{noncomphopfs2nd})) to construct an $Sp(4; \mathbb{R})\simeq SO(2,3)$ extension of  squeezed state  \cite{Hasebe-2019}. 
In the present paper, we propose an $SO(4,1)$ version of squeezed state based on the $H^4$ geometry (the right of (\ref{noncomphopfs2nd})). 
It turns out that the $SO(4,1)$ squeezed vacuum attains a more natural generalization of the original squeezed vacuum  than the previous $SO(2,3)$ case in several respects.  The $SO(4,1)$ squeezed vacuum accommodates an interesting  spin structure  related to  the hierarchy of  
the non-compact 2nd  Hopf map: 
\begin{center}
\begin{tabular}{cccccc}
\\ 
 & & $S^{3}$ &   $\overset{S^{1}}\longrightarrow $ & $S^{2}$ &  \\ 
 &  $H^{4,3}$ & $\longrightarrow$ & $H^{4}$ &  &  
\end{tabular}
\end{center}
$H^{4,3}$ can be regarded as a fibre-bundle of base-manifold $H^{4}$ with $S^3$-fibre which itself  denotes the  fibre-bundle space of  1st Hopf map of the Bloch sphere. This  implies the existence of inherent spin geometry in the present $SO(4,1)$  formulation.

Before proceeding to detail discussions,   we mention  differences between the previous $SO(2,3)$ formulation \cite{Hasebe-2019} and the present $SO(4,1)$ formulation in 
 a group theory point of view.    
 For the original $SO(2,1)$ squeezed state, the associated two-hyperboloid is given by  
\begin{align}
H^{2}~ &\simeq ~
SO(2,1)/SO(2)~ \simeq ~SU(1,1)/U(1)~\simeq~Sp(2; \mathbb{R})/U(1)\nn\\
&~ \simeq ~U(1; \mathbb{H}')/U(1).   
\label{h2so21}
\end{align}
The symmetry group  $SO(2,1)$ allows  Majorana representation as well as the Dirac representation, which  leads to  one-mode  and two-mode squeeze operators. 
The holonomy group  is a compact group $SO(2)\simeq U(1)$.  
Meanwhile in the previous work  \cite{Hasebe-2019}, we utilized a Bloch four-hyperboloid (right of (\ref{noncomphopfs2nd})):
\begin{align}
H^{2,2} &~\simeq 
~SO(2,3)/SO(2,2)~ \simeq ~Sp(4; \mathbb{R})/(SU(1,1)\otimes SU(1,1)))\nn\\
&~ \simeq ~U(2; \mathbb{H}')/(U(1; \mathbb{H}')\otimes U(1; \mathbb{H}')). 
\end{align}
The symmetry group $SO(2,3)$ also 
accommodates  Majorana representation as well as the Dirac representation, 
and the corresponding two-mode and four-mode $SO(2,3)$ squeezed states are constructed.   
A crucial difference to the original case is  the holonomy group $SO(2,2) \simeq SU(1,1)\otimes SU(1,1)$, which is non-compact unlike $SO(2)$. 
In the present work, we will adopt the other Bloch four-hyperboloid (left of (\ref{noncomphopfs2nd})): 
\begin{align}
H^{4} &~\simeq 
~SO(4,1)/SO(4)~ \simeq ~USp(2,2)/(SU(2)\otimes SU(2)))\nn\\
&~ \simeq ~U(1, 1; \mathbb{H})/(U(1; \mathbb{H})\otimes U(1; \mathbb{H})). 
\end{align}
The symmetry group $SO(4,1)$ 
does not accommodate  Majorana representation,  and so the two-mode realization of  $SO(4,1)$  squeezed state is not possible, and hence  the $SO(4,1)$ squeezed operator can be realized only by a four-mode Dirac representation.  
An analogous point to the original case  is 
the holonomy group, which is given by a compact group  $SO(4) \simeq SU(2)\otimes SU(2)$  like $SO(2)$.

This paper is organized as follows. 
In Sec.\ref{sec:blochsphblochhyp}, we review spin-coherent state and  squeezed state with emphasis on their relations to the geometry of  Bloch sphere and hyperboloid. 
We also discusses interesting relationship between spin-coherent states and squeezed states. In Sec.\ref{sec:blochfour}, the $SO(4,1)$ squeeze operator is introduced based on the geometry of the Bloch four-hyperboloid $H^4$. In Sec.\ref{sec:so41sqvac}, we propose a physical interpretation of the $SO(4,1)$ squeezed vacuum in spin geometry and explore its basic  properties, such as  von Neumann entropy, spin correlations and uncertainty relations.    
Sec.\ref{sec:summary} is devoted to summary and discussions.

\section{Bloch hyperboloid and  squeezed states}\label{sec:blochsphblochhyp}

 We begin with a review of the Bloch sphere and Bloch hyperboloid and their associated  coherent states.  The Schwinger operator formulation is not only useful to represent  spin state with arbitrary spin  but  also closely  
 related to  the squeeze operator.

\subsection{Bloch sphere and spin-coherent state}

\begin{figure}[tbph]
\center
\includegraphics*[width=110mm]{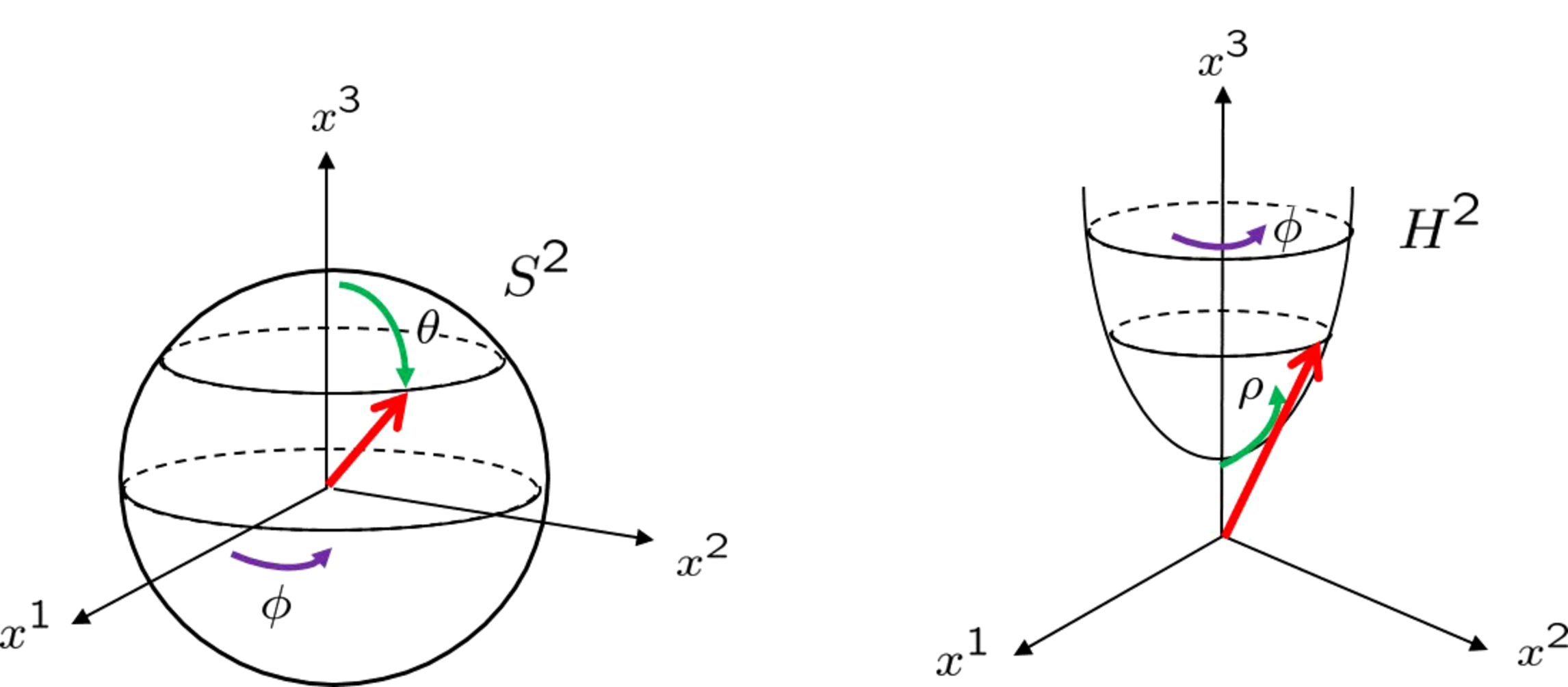}
\caption{Bloch sphere and Bloch hyperboloid for the spin and pseudo-spin-coherent states.   }
\label{s2h2.fig}
\end{figure}

The geometry of the Bloch sphere \cite{Bloch-1946} stems from the 1st Hopf map:\footnote{See \cite{Mosseri-Dandoloff-2001} for the roles of the 2nd Hopf map in the context of entanglement of qubits.  } 
\be
S^2 ~\overset{S^1}{\longrightarrow}~S^2,  \label{hopfmap1comp}
\ee
and the Hopf map is explicitly realized as the map from the 2-component normalized spinor  $(\psi^{\dagger}\psi=1:S^3)$ to a normalized three component vector $(\sum_{i=1}^3 x_ix_i=1:S^2)$ \cite{Hasebe-2010}: 
\be
\psi=\begin{pmatrix}
\psi_1 \\
\psi_2
\end{pmatrix} ~~\rightarrow~~x_i=\psi^{\dagger}\sigma_i\psi. 
\ee
Here, $\sigma_i$ $(i=1,2,3)$ are the Pauli matrices: 
\be
\sigma_x =\begin{pmatrix}
0 & 1 \\
1 & 0 
\end{pmatrix}, ~~\sigma_y =\begin{pmatrix}
0 & -i \\
i & 0 
\end{pmatrix}, ~~\sigma_z =\begin{pmatrix}
1 & 0  \\
0 & -1 
\end{pmatrix}. 
\ee
When we parameterize $x_i ~\in~S^2$ by using the usual polar coordinates 
\be
x_1=\sin\theta\cos\phi,~~ x_2=\sin\theta\sin\phi,~~ x_3=\cos\theta, 
\ee
$\psi$ can be expressed as 
\be
\psi(\phi,\theta,\chi)= \begin{pmatrix}
\cos(\frac{\theta}{2}) ~e^{-i\frac{1}{2}\phi} \\
\sin(\frac{\theta}{2}) ~ e^{i\frac{1}{2}\phi}
\end{pmatrix}e^{-i\frac{1}{2}\chi}. \label{hopfspcomp1}
\ee
Here, $e^{-i\frac{1}{2}\chi}$ denotes the overall $U(1)$ phase that corresponds to the $S^1$ fibre in (\ref{hopfmap1comp}). Importantly, $\psi$ has a physical meaning as spin-coherent state \cite{Arecchi-Courtens-Gilmore-Thomas-1972,Perelomov-1972,Radcliffe-1971} that is aligned to the direction of $(\theta, \phi)$ on the $S^2$ (see the left of Fig.\ref{s2h2.fig}): 
\be
(\bs{x}(\theta, \phi) \cdot \bs{\sigma} )  \psi =\psi. \label{eqspincoh12}
\ee
Applying the Euler angle rotation \cite{Sakurai-book-1993} to the  spin state aligned to  $z$ direction, we can readily reproduce (\ref{hopfspcomp1}) as 
\be
\psi(\phi,\theta,\chi)\equiv  
D^{(1/2)}(\phi,\theta,\chi)\begin{pmatrix}
1 \\
0 
\end{pmatrix} =\cos(\frac{\theta}{2}) ~e^{-i\frac{1}{2}(\chi+\phi)}\begin{pmatrix}
1 \\
0\end{pmatrix} + \sin(\frac{\theta}{2}) ~e^{-i\frac{1}{2}(\chi-\phi)}\begin{pmatrix}
0 \\
1\end{pmatrix}.  \label{euleranrot}
\ee
The rotation matrix in (\ref{euleranrot}) is known as the  Wigner's (spin $1/2$) D-matrix  of the $SU(2)$ group : 
\be
D^{(1/2)}(\phi,\theta,\chi) =e^{-i\frac{\phi}{2} \sigma_z}e^{-i\frac{\theta}{2} \sigma_y}e^{-i\frac{\chi}{2} \sigma_z}  
=
\begin{pmatrix}
\cos(\frac{\theta}{2}) ~e^{-i\frac{1}{2}(\chi+\phi)} &  -\sin(\frac{\theta}{2}) ~e^{i\frac{1}{2}(\chi-\phi)} \\
\sin(\frac{\theta}{2}) ~e^{-i\frac{1}{2}(\chi-\phi)} & \cos(\frac{\theta}{2}) ~e^{i\frac{1}{2}(\chi+\phi)}
\end{pmatrix}, 
\ee
where the range of the parameters are 
\be
0\le \phi < 2\pi, ~~0\le \theta < \pi, ~~0\le \chi < 4\pi.  
\ee

For later convenience, we also introduce  coset representative of the $SU(2)$ group.\footnote{Coset representative (see \cite{Gilmore-2008} for instance) is also referred to as the non-linear realization \cite{Coleman-Wess-Zumino-1969,CallanJr-Coleman-Wess-Zumino-1969,Salam-Strathdee-1982}.}   
As $S^2$ is expressed as a coset 
\be
S^2 \simeq S^3/S^1 \simeq SU(2)/U(1),  
\ee
the corresponding coset representative  is obtained as   
\be
e^{i\theta \sum_{m=1,2}{y_{m}\sigma_{m 3}}} 
=e^{-\frac{\theta}{2}(e^{-i\phi}~\sigma^+-e^{i\phi}~\sigma^-)}=\begin{pmatrix}
\cos\frac{\theta}{2} & -\sin\frac{\theta}{2}~e^{-i\phi} \\
\sin\frac{\theta}{2}~e^{i\phi} & \cos\frac{\theta}{2}
\end{pmatrix}, \label{su2costrepfors2}
\ee
where 
\be
{y}_1=\cos\phi,~~{y}_2=\sin\phi, 
\ee
and  
\be
\sigma_{ij}=\frac{1}{2}\epsilon_{ijk}\sigma_k , ~~~~\sigma^+ =\begin{pmatrix}
0 & 1 \\
0 & 0 
\end{pmatrix},~~\sigma^- =\begin{pmatrix}
0 & 0 \\
1 & 0 
\end{pmatrix} .
\ee
(\ref{su2costrepfors2}) can be factorized as 
\be 
\text{Euler decomposition}:~e^{i\theta \sum_{m=1,2}{y_{m}\sigma_{m 3}}} =  e^{-i\frac{\phi}{2} \sigma_z}  {e^{-i\frac{\theta}{2} \sigma_y}} e^{i\frac{\phi}{2} \sigma_z}=D^{(1/2)}(\phi,\theta,-\phi)  \label{eulerdecomsu2}
\ee 
and 
\be 
\text{Gauss decomposition}:~ e^{i\theta \sum_{m=1,2}{y}_{m}\sigma_{m 3}} =e^{-\tan\frac{\theta}{2}e^{-i\phi}\sigma^+} e^{-\ln (\cos\frac{\theta}{2})\sigma_z} e^{\tan\frac{\theta}{2} e^{i\phi}\sigma^-}.  
\label{gauss-decomp-su2}
\ee 
Notice that the Euler decomposition (\ref{eulerdecomsu2}) is realized as a special case of the D-matrix and the physical meaning of the coset representative is transparent  in this decomposition.

\subsubsection{Schwinger operator realization}

\begin{figure}[tbph]
\center
\includegraphics*[width=45mm]{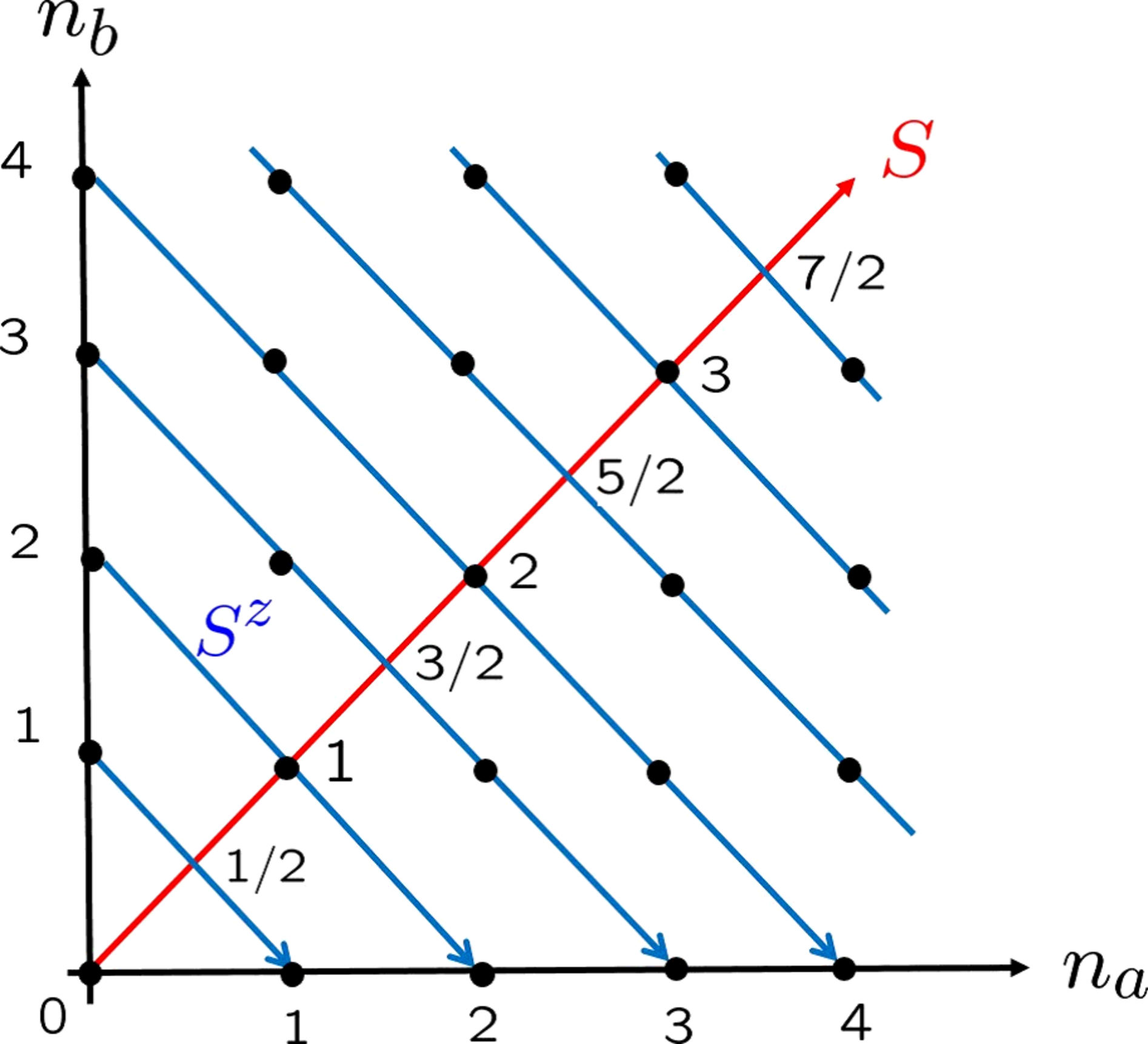}
\caption{The $SU(2)$ irreducible representation with fixed $S$ is specified by each of the oblique blue lines of $n_a+n_b=2S$ .}
\label{sum.fig}
\end{figure}

We construct the spin-coherent state with arbitrary spin magnitude using the Schwinger operator formalism \cite{Schwinger-1965,Sakurai-book-1993}.  
With  the Schwinger boson operator  
\be
\hat{\psi}\equiv \begin{pmatrix}
a \\
b
\end{pmatrix} \label{su2psischw}
\ee
subject to 
\be
[a, a^{\dagger}]=[b,b^{\dagger}]=1, ~~~[a,b]=[a,b^{\dagger}]=0, \label{sommureaandb}
\ee
the spin operators are expressed as 
\be
\hat{S}_i=\hat{\psi}^{\dagger}\frac{1}{2}\sigma_i \hat{\psi} . 
\ee
In detail, 
\be
\hat{S}_x=\frac{1}{2}(a^{\dagger}b +b^{\dagger}a), ~~~~\hat{S}_y=-i\frac{1}{2}(a^{\dagger}b-b^{\dagger}a),~~~~\hat{S}_z=\frac{1}{2}(a^{\dagger}a-b^{\dagger}b), 
\ee
and so 
\be 
{\hat{\bs{S}}}^2  =\hat{S}(\hat{S}+1)
\ee 
with 
\be
\hat{S}=\frac{1}{2}(\hat{n}_a+\hat{n}_b)=\frac{1}{2}(a^{\dagger}a+b^{\dagger}b). 
\ee
The spin states $|S, S_z\rangle$ that satisfy 
\be
{\hat{\bs{S}}}^2 |S, S_z\rangle = S(S+1)|S, S_z\rangle , ~~~~~\hat{S}_z|S, S_z\rangle =S_z|S, S_z\rangle  
\ee
 are constructed as 
\be
|S, S_z\rangle \equiv |n_a, n_b\rangle =\frac{1}{\sqrt{n_a!n_b!}} ~(a^{\dagger})^{n_a}(b^{\dagger})^{n_b}|0,0\rangle,  \label{angunnequiv}
\ee
where the relations between the boson numbers $(n_a, n_b)$ and spin values $(S, S_z)$ are depicted in Fig.\ref{sum.fig}, and  
hereafter we utilize $|n_a, n_b\rangle$ and $|S, S_z\rangle$ interchangeably. 

We introduce D-operator as  
\be
\hat{D}(\phi, \theta, \chi) \equiv e^{-i\phi \hat{S}_z} e^{-i\theta \hat{S}_y} e^{-i\chi \hat{S}_z} 
\ee
whose matrix representation is the D-matrix:   
\be
\langle S, S_z|\hat{D}(\phi, \theta, \chi)  |S', S'_z\rangle =  {D}^{(S)}(\phi, \theta, \chi)_{S_z, S'_z} \cdot \delta_{S, S'},   \label{dmatop}
\ee
$i.e.$,  
\be
{D}^{(S)}(\phi, \theta, \chi) =e^{-i\phi S_z} e^{-i\theta S_y}e^{-i\chi S_z}. 
\label{defwigfuncmat}
\ee
In (\ref{defwigfuncmat}), $S_{i=x,y,z}$ denote the $SU(2)$ matrices with spin magnitude $S$. 
Under the $SU(2)$ transformation of $\hat{D}(\phi, \theta, \chi)$,  (\ref{su2psischw}) transforms as an $SU(2)$ spinor 
\be
\hat{D}(\phi, \theta, \chi)^{\dagger}~\hat{\psi}~ \hat{D}(\phi, \theta, \chi) ={D}^{(1/2)}(\phi, \theta, \chi)~\hat{\psi}.
\ee
In a similar manner to (\ref{euleranrot}), we can readily construct the spin-coherent state with arbitrary spin magnitude:
\be
|S, S_z(\phi,\theta,\chi)\rangle =\hat{D}(\phi,\theta,\chi)|S,S_z\rangle =e^{-iS_z\chi } \hat{D}(\phi,\theta,0)|S,S_z\rangle,  
\label{spincohschw}
\ee
which satisfies 
\begin{align}
&{\hat{\bs{S}}}^2 |S , S_z( \phi, \theta, \chi)\rangle  =\frac{1}{4}(n_a+n_b) (n_a+n_b+2) |S , S_z(\phi, \theta, \chi)\rangle , \nn\\
&(\bs{x}(\theta, \phi)\cdot \hat{\bs{S}}) |S , S_z(\phi, \theta, \chi)\rangle  =\frac{1}{2}(n_a-n_b)  |S , S_z(\phi, \theta, \chi)\rangle. 
\end{align}
With (\ref{dmatop}), (\ref{spincohschw}) can be concisely expressed as a linear combination of the spin states with expansion coefficients of the components of D-matrix:\footnote{Interestingly, D-matrices themselves have a physical meaning; $D^{(S)}(\phi,\theta, \chi)_{S'_z, S_z}$ denote  $2S+1$ degenerate  eigenstates of  $(S-|S_z|)$th Landau level in the monopole  background with magnetic charge $S_z$ (see \cite{Hasebe-2015} for instance). }   
\be 
|S, S_z(\phi,\theta,\chi)\rangle =\sum_{S'_z=-S}^{S} D^{(S)}(\phi,\theta, \chi)_{S'_z, S_z}~|S, S'_z\rangle  ,  \label{spincohexpa}
\ee
which is a ket state expression of (\ref{euleranrot}) generalizing the spin $1/2$ to arbitrary spin magnitude.    
In the Schwinger boson notation, (\ref{spincohexpa}) is represented as 
\be
|S , S_z(\phi, \theta, \chi)\rangle 
=\sum_{n'_a,n'_b =0 }^{2S}  D^{(S=\frac{1}{2}(n_a+n_b))}(\phi, \theta, \chi)_{\frac{1}{2}(n'_a-n'_b), \frac{1}{2}(n_a-n_b)}  |n'_a, n'_b\rangle. 
\ee

\subsection{Bloch  hyperboloid and pseudo-spin-coherent state}

Here, we extend the above discussions  to the case of hyperboloid. 
The  $SU(1,1)$ version of  D-operator realizes the squeeze operator 
 in its  special form.

\subsubsection{$SU(1,1)$ algebra and the Schwinger operator realization  }

The non-compact Hopf map is given by 
\be
H^{2,1} ~~\overset{S^1}\longrightarrow ~~ H^{2}, 
\ee
and, in the same sense of the Bloch sphere, we refer to the base-manifold $H^2$ as the Bloch hyperboloid.  $H^{2,1}$ is a group manifold  of the $SU(1,1)$ whose invariant matrix is $\sigma_z$. 
The $SU(1,1)$ matrix generators can be realized as  the following non-Hermitian matrices
\be
\frac{1}{2}\tau^i =\{i\frac{1}{2}\sigma_x, i\frac{1}{2}\sigma_y, \frac{1}{2}\sigma_z \}, 
\ee
which satisfy 
\be
\{\tau^i, \tau^j\} =2\eta^{ij}, ~~~[\tau^i, \tau^j] =-2i\epsilon^{ijk}\tau_k, 
\ee
with 
\be
\eta_{ij}=\text{diag}(- , -, +), ~~\epsilon^{123}=1. 
\ee
With two-component spinor subject to $\psi^{\dagger}\sigma_z \psi=|\psi_1|^2-|\psi_2|^2 = 1: H^{2,1}$, the non-compact Hopf map is explicitly realized as 
\be
\psi ~~\rightarrow ~~x^i=\psi^{\dagger}\kappa^i \psi, \label{noncomp1st}
\ee
where $\kappa^i$ are Hermitian matrices made of $\tau^i$: 
\be
\kappa^i=\sigma_z\tau^i= \{-\sigma_y, \sigma_x, 1_2 \}. 
\ee
$x^i$ (\ref{noncomp1st}) satisfy 
\be
\eta_{ij}x^ix^j=x^ix_i=-{x^1}x^1-{x^2}x^2+{x^3}x^3=(\psi^{\dagger}\kappa\psi)^2=1, 
\ee
and denote the coordinates on  $H^2$. 
When $x^i$ are parameterized as 
\be
x^1  =\sinh\rho \sin\phi ,~~ x^2=\sinh\rho\cos\phi ,~~ x^3=\cosh\rho,  
\ee
$\psi$ can be  realized  as 
\be
\psi(\phi,\rho,\chi) =
\begin{pmatrix}
\cosh(\frac{\rho}{2})~ e^{i\frac{\phi}{2}} \\
\sinh(\frac{\rho}{2}) ~e^{-i\frac{\phi}{2}} 
\end{pmatrix}e^{i\frac{\chi}{2}},   \label{noncomphopf}
\ee
which also acts as the pseudo-spin-coherent state \cite{Perelomov-1972, Gilmore-Yuan-1987,Combescure-Robert-book-2012} (see the right of Fig.\ref{s2h2.fig}): 
\be
(x^i(\rho,\phi) ~\tau_i) \psi =\psi.  
\ee
$\psi$ can be derived by acting the $SU(1,1)$ rotation to the pseudo-spin state aligned to   $z$-direction:
\be
\psi(\phi,\rho,\chi) ={D}^{(1/2)}(\phi, \rho, \chi) \begin{pmatrix}
1 \\
0
\end{pmatrix} , 
\ee
where ${D}^{(1/2)}(\phi, \rho, \chi)$  is   the $SU(1,1)$ version of the D-matrix\footnote{
One may explicitly verify that (\ref{su11dmat})  satisfies  the conditions of the $SU(1,1)$ group: 
\be
{D}^{(1/2)}(\phi, \rho, \chi)^{\dagger} ~\sigma_z~ D^{(1/2)}(\phi, \rho, \chi) =\sigma_z, ~~\det(D^{(1/2)}(\phi, \rho, \chi))=1. 
\ee
} 
\be
{D}^{(1/2)}(\phi, \rho, \chi) = e^{i\frac{\phi}{2}\tau^z}e^{-i\frac{\rho}{2}\tau^x} e^{i\frac{\chi}{2}\tau^z} =\begin{pmatrix}
\cosh (\frac{\rho}{2}) ~e^{i\frac{1}{2}(\chi+\phi)} & \sinh (\frac{\rho}{2}) ~e^{-i\frac{1}{2}(\chi-\phi)} \\
\sinh (\frac{\rho}{2})~ e^{i\frac{1}{2}(\chi-\phi)} &  \cosh (\frac{\rho}{2}) ~e^{-i\frac{1}{2}(\chi+\phi)} 
 \end{pmatrix}. \label{su11dmat}
\ee
The ranges of the parameters are  
\be
\rho =[0, \infty), ~~\phi=[0, 2\pi), ~~\chi=[0, 4\pi). 
\ee

With the Schwinger boson operators, we can construct the $SU(1,1)$ operators as  
\be
\hat{T}^i =\hat{\psi}^{\dagger}\kappa^i \hat{\psi}.  \label{defogti}
\ee
Here, $\hat{\psi}$ is defined as 
\be
\hat{\psi} \equiv \begin{pmatrix}
a \\
b^{\dagger}
\end{pmatrix}, 
\ee
whose components satisfy 
\be
[\hat{\psi}_{\alpha}, \hat{\psi}_{\beta}^{\dagger}] =(\sigma_z)_{\alpha\beta}, ~~~[\hat{\psi}_{\alpha}, \hat{\psi}_{\beta}]=0.
\ee
The $SU(1,1)$ Hermitian operators  are represented as \cite{Wodkiewicz-Eberly-1985, Yurke-McCall-Klauder-1986, Gerry-1987}
\be
\hat{T}^x =i\frac{1}{2}(a^{\dagger}b^{\dagger}-ab) ,~~\hat{T}^y =\frac{1}{2}(a^{\dagger}b^{\dagger}+ab) ,~~\hat{T}^z=\frac{1}{2}(a^{\dagger}a+b^{\dagger}b)+\frac{1}{2}, 
\ee
and the ladder operators are 
\be
\hat{T}^+=a^{\dagger}b^{\dagger}, ~~~\hat{T}^-=ab. 
\ee
The $SO(2,1)$ Casimir is derived as 
\be
C=\hat{T}_i\hat{T}^i=-(\hat{T}^x)^2 -(\hat{T}^y)^2 +(\hat{T}^z)^2  =\frac{1}{4}({\hat{\psi}}^{\dagger}\sigma_z\hat{\psi})~({\hat{\psi}}^{\dagger}\sigma_z\hat{\psi} +2),
\ee
where 
\be
{\hat{\psi}}^{\dagger} \sigma_z\hat{\psi}=a^{\dagger}a-bb^{\dagger}=a^{\dagger}a-b^{\dagger}b-1. 
\ee
The $SU(1,1)$ D-operator is given by 
\be
\hat{{D}}(\phi, \rho, \chi) = e^{i{\phi}\hat{T}^z}e^{-i{\rho} \hat{T}^x} e^{i{\chi}\hat{T}^z}.
\ee
Under the $SU(1,1)$ transformation of $\hat{D}(\phi, \rho, \chi)$,  $\hat{\psi}$ transforms as an $SU(1,1)$ spinor 
\be
\hat{{D}}(\phi, \rho, \chi)^{\dagger}~\hat{\psi}~ \hat{{D}}(\phi, \rho, \chi) ={{D}}^{(1/2)}(\phi, \rho, \chi)~\hat{\psi}.
\ee

\subsubsection{$SU(1,1)$ coset representative and the squeeze operator}

In the parameterization 
\be
y_1(\phi)=\sin\phi, ~~y_2(\phi)=\cos\phi ~\in~S^1, 
\ee
an $SU(1,1)$ coset representative  for $H^2\simeq SU(1,1)/U(1)$  is obtained as 
\be 
M(\rho, \phi) =e^{i\frac{\rho}{2}\epsilon_{mn}y^m(\phi)~\tau^n}
 =e^{-\frac{\rho}{2}e^{i\phi} t^+ +\frac{\rho}{2}e^{-i\phi} t^-}
=\begin{pmatrix}
\cosh(\frac{\rho}{2}) & -\sinh(\frac{\rho}{2})~e^{i\phi} \\
-\sinh(\frac{\rho}{2})~e^{-i\phi} & \cosh(\frac{\rho}{2}) 
\end{pmatrix},   \label{cosetrepresu11}
\ee 
with $\epsilon_{12}=-\epsilon_{21}=1$ and 
\be
t^+=-i\frac{1}{2}\tau^x+\frac{1}{2}\tau^y=\begin{pmatrix}
0 & 1 \\ 
0 & 0 
\end{pmatrix}, ~~t^-=i\frac{1}{2}\tau^x+\frac{1}{2}\tau^y=\begin{pmatrix}
0 & 0 \\ 
-1 & 0 
\end{pmatrix}. 
\ee
(\ref{cosetrepresu11}) can be factorized as 
\be
\text{Euler decomposition}~:~M(\rho, \phi) = e^{i\frac{\phi}{2} \tau^z} e^{i\frac{\rho}{2} \tau^x} e^{-i\frac{\phi}{2} \tau^z}={D}^{(1/2)}(\phi, -\rho, -\phi), \label{euler-decomp-sp2r} 
\ee
and 
\begin{align}  
\text{Gauss decomposition}~:~&M(\rho, \phi) =e^{-\tanh (\frac{\rho}{2}) ~e^{i\phi} ~t^+}~ e^{-\ln(\cosh \frac{\rho}{2}) \tau^z} e^{\tanh (\frac{\rho}{2}) ~e^{-i\phi}~ t^-}\nn\\
&~~~~~~~~~~= e^{-\tanh\frac{\rho}{2} y^m  (\frac{1}{2} \tau_m-i\tau_{m3}) }~ e^{-\log(\cosh\frac{\rho}{2})~ \tau_z} ~e^{\tanh\frac{\rho}{2} y^m  (\frac{1}{2} \tau_m+i\tau_{m3}) },   \label{gauss-decomp-sp2r}
\end{align} 
where 
\be
\tau^{ij} =-i\frac{1}{4}[\tau^i, \tau^j]=-\frac{1}{2}\epsilon^{ijk}\tau_k. 
\ee 
In the Gauss decomposition (\ref{gauss-decomp-sp2r}), the Poincar\'e Disc coordinates, $\tanh \frac{\rho}{2} ~y^m(\phi)$, appear on the shoulder of the exponent. 
The operator that corresponds to the coset representative (\ref{cosetrepresu11}) is constructed as 
\be 
\hat{S}(\rho, \phi) =e^{i\frac{\rho}{2}\epsilon_{mn}y^m(\phi)~\hat{T}^n }=\exp\biggl(-\frac{\rho}{2}~e^{i\phi}~a^{\dagger}b^{\dagger}+ \frac{\rho}{2}~e^{-i\phi}~ ab \biggr).  \label{squeezeoriop}
\ee 
Notice that (\ref{squeezeoriop}) is nothing but the two-mode squeeze operator in quantum optics \cite{Hollenhorst-1979,Caves-1981,Walls-1983}.  Thus, the $SU(1,1)$ D-operator realizes the squeeze operator in its special form.  Since the $SU(1,1)$ D-operator signifies hyperbolic rotation, the squeeze operator can be interpreted as  a hyperbolic rotation operator.  
As exemplified in the above, 
(\ref{squeezeoriop}) is factorized as 
\begin{align} 
\text{Euler decomposition}~:~
\hat{S}(\rho, \phi)& =\hat{D}(\phi, -\rho, -\phi) =e^{i\phi \hat{T}^z} e^{i\rho \hat{T}^x} e^{-i\phi \hat{T}^z}\nn\\
&=e^{i\frac{\phi}{2}(\hat{n}_a+\hat{n}_b)}e^{-\frac{\rho}{2}(a^{\dagger}b^{\dagger}+ab)} e^{-i\frac{\phi}{2}(\hat{n}_a+\hat{n}_b)},  
\label{eulerdecsu11}
\end{align} 
and 
\begin{align}
\text{Gauss decomposition}~:~\hat{S}(\rho, \phi)& 
=e^{-\tanh\frac{\rho}{2}e^{i\phi} ~\hat{T}^+} e^{-2\ln(\cosh \frac{\rho}{2}) \hat{T}^z} e^{\tanh\frac{\rho}{2}e^{-i\phi}~\hat{T}^-}\nn\\
&=\frac{1}{\cosh\frac{\rho}{2}}~ e^{-\tanh(\frac{\rho}{2})e^{i\phi} ~ a^{\dagger}b^{\dagger}}~
\biggl(\frac{1}{\cosh (\frac{\rho}{2})  }\biggr)^{\hat{n}_a+\hat{n}_b} ~e^{\tanh(\frac{\rho}{2})e^{-i\phi}~ ab}. 
\label{twomodedisq}
\end{align}
The Euler decomposition (\ref{eulerdecsu11}) demonstrates the physical meaning of the squeeze operation, $i.e.$, the $SU(1,1)$ hyperbolic rotation. 
Meanwhile, the Gauss decomposition  is very useful in deriving the number state expansion of squeezed state \cite{Gerry-1987}, as   (\ref{twomodedisq}) is realized as an almost  normal ordered form. 

The phase $\phi$ 
can be absorbed in 
the  phase redefinition of the Schwinger boson operators\footnote{
The commutation relations (\ref{sommureaandb}) are immune to the phase redefinition of each Schwinger operator (\ref{redefphaab}). 
}   
\be
a ~~\rightarrow ~~e^{-i\alpha}~a, ~~~~b ~~\rightarrow ~~e^{-i\beta}~b, \label{redefphaab}
\ee
the squeeze operator is transformed as 
\be
\hat{S}(\rho, \phi)=\exp\biggl(-\frac{\rho}{2}~e^{i\phi}~a^{\dagger}b^{\dagger}+ \frac{\rho}{2}~e^{-i\phi}~ ab \biggr)~~\rightarrow~~S(\rho)=\exp\biggl(-\frac{\rho}{2}~a^{\dagger}b^{\dagger}+ \frac{\rho}{2}~ ab \biggr),  
\ee
where
\be
\alpha=\frac{1}{2}\phi+\gamma,~~~\beta=\frac{1}{2}\phi-\gamma.
\ee
(The degree of freedom of $\gamma$ still remains.)  

\subsection{Two-mode $SO(2,1)$ squeeze vacuum and entangled spin-coherent states}\label{sec:twomodesq}

\subsubsection{$SO(2,1)$ two-mode squeeze operator and the interaction Hamiltonian}

With the Gauss decomposition (\ref{twomodedisq}), 
the two-mode squeezed vacuum\footnote{See Appendix \ref{append:basictwomodevac} about  basic properties of the two-mode squeezed vacuum.}   is readily obtained as  
\be
|\text{sq} (\rho, \phi)\rangle =\hat{S}(\rho, \phi)|0\rangle \otimes |0\rangle 
=\frac{1}{\cosh \frac{\rho}{2}} \sum_{n=0}^{\infty}~\biggl(-\tanh
(\frac{\rho}{2})\biggr)^n~e^{i n \phi}~|n\rangle \otimes |n\rangle,  \label{numberstaterepsq2r}
\ee 
where 
\be
|n\rangle \otimes |n\rangle =\frac{1}{\sqrt{n!}}{a^{\dagger}}^n|0\rangle \otimes \frac{1}{\sqrt{n!}} {b^{\dagger}}^n|0\rangle. 
\ee
$|\text{sq} (\rho, \phi)\rangle$ signifies an entangled state of the number states. 
Removing the right $U(1)$ factor of (\ref{eulerdecsu11}), we introduce the Schwinger-type squeeze operator as   \cite{Hasebe-2019} 
\be
\hat{\mathcal{S}}(\rho, \phi) =e^{i\phi \hat{T}^z} e^{i\rho \hat{T}^x} . 
\ee 
The difference between the squeezed vacuum of the Dirac-type and that of the Schwinger-type is just the phase factor: 
\be
|\text{sq} (\rho, \phi)\rangle =\hat{S}(\rho, \phi)|0\rangle \otimes |0\rangle = e^{-i\frac{\phi}{2}}~\hat{\mathcal{S}}(\rho, \phi)|0\rangle\otimes |0\rangle ,   \label{sqsqdiff}
\ee
and so the Dirac-type and Schwinger-type squeezed vacua are physically identical. 
In the interaction picture, the two-mode Hamiltonian  realizes as the time-evolution operator of the squeeze operator (\ref{squeezeoriop}), which is 
\be
\hat{H}^{\text{int}}(\rho, \phi) =-i\frac{\rho}{2} e^{i\phi} a^{\dagger}b^{\dagger} + i \frac{\rho}{2} e^{-i\phi} ab. 
\ee

\subsubsection{Two two-mode squeezed vacua as entangled spin-coherent states} 

For later convenience,  we point out an interesting correspondence between the $SU(2)$ spin-coherent states and the $SU(1,1)$ squeezed states.   Let us consider  a direct product of two independent squeezed vacua with same squeeze parameter $\rho$. We also assume that  
  a squeezed vacuum is realized in  the Schwinger boson space of $a$ and $c$, and another  in the Schwinger boson space of $b$ and $d$:\footnote{(\ref{sqsq}) is a four-mode analogue of the two-mode $SU(1,1)\otimes SU(1,1)$ state of \cite{Gerry-Benmoussa-2000}.} 
\begin{align}
|\text{sq} (\rho, \phi)\rangle_{a,c} \otimes   |\text{sq} (\rho,\phi')\rangle_{b,d}&=\frac{1}{\cosh^2\frac{\rho}{2}}~e^{-\frac{\rho}{2}  e^{i\phi} ~a^{\dagger}c^{\dagger} }
 e^{-\frac{\rho}{2}  e^{i\phi'}b^{\dagger}d^{\dagger}}~ |0,0,0,0\rangle \nn\\
& =\frac{1}{\cosh^2\frac{\rho}{2}}\sum_{n,m =0}^{\infty} (-\tanh(\frac{\rho}{2}))^{n+m} e^{i(n\phi+m\phi')}|n, m, n, m\rangle, \label{sqsq}
\end{align}
where $|n,m,n,m \rangle \equiv |n\rangle_a |m\rangle_b |n\rangle_c |m\rangle_d$.  
Recall that the Schwinger interpretation of spin states, in which  $|n,m\rangle$ is interpreted as the spin state $|S, S_z\rangle$, and  the sum of all   particles numbers   is replaced with the sum of all possible spin degrees of freedom (see Fig.\ref{sum.fig}): 
\be
\sum_{n,m=0,1,2,3,\cdots} ~~\rightarrow~~\sum_{S=0,1/2,1,3/2,\cdots} ~\sum_{S_z=-S}^S. \label{changenmtossz}
\ee
With this replacement, (\ref{sqsq}) can be rewritten as 
\be
|\text{sq} (\rho, \phi)\rangle_{a,c} \otimes   |\text{sq} (\rho,\phi')\rangle_{b,d}  =\frac{1}{\cosh^2(\frac{\rho}{2})}\sum_{S =0, 1/2, 1, \cdots }^{\infty}  (-\tanh(\frac{\rho}{2}))^{2S} e^{iS(\phi+\phi')} \sum_{S_z=-S}^S e^{iS_z(\phi-\phi')}~|S, S_z\rangle_{a,b}\otimes  |S, S_z\rangle_{c,d},  \label{prodtwomodesqussss}
\ee 
which denotes a linear combination of  direct products of two identical spin states with all possible  spins.  
In each spin sector, we have a   maximally entangled spin state: 
\be
\sum_{S_z=-S}^S e^{iS_z(\phi-\phi')}~|S, S_z\rangle_{a,b}\otimes  |S, S_z\rangle_{c,d}. 
\ee
Though  in (\ref{prodtwomodesqussss}) the left and right-hand sides   are  mathematically identical thorough the Schwinger operator formalism, they   describe totally different physics: The left-hand side denotes a 
 $\it{separable~state}$ of two squeezed states, while   the right-hand side stands for a  $\it{maximally~entangled}$ spin state.

\section{Bloch four-hyperboloid and  $SO(4,1)$ squeeze operator}\label{sec:blochfour}

\begin{figure}[tbph]
\center
\includegraphics*[width=160mm]{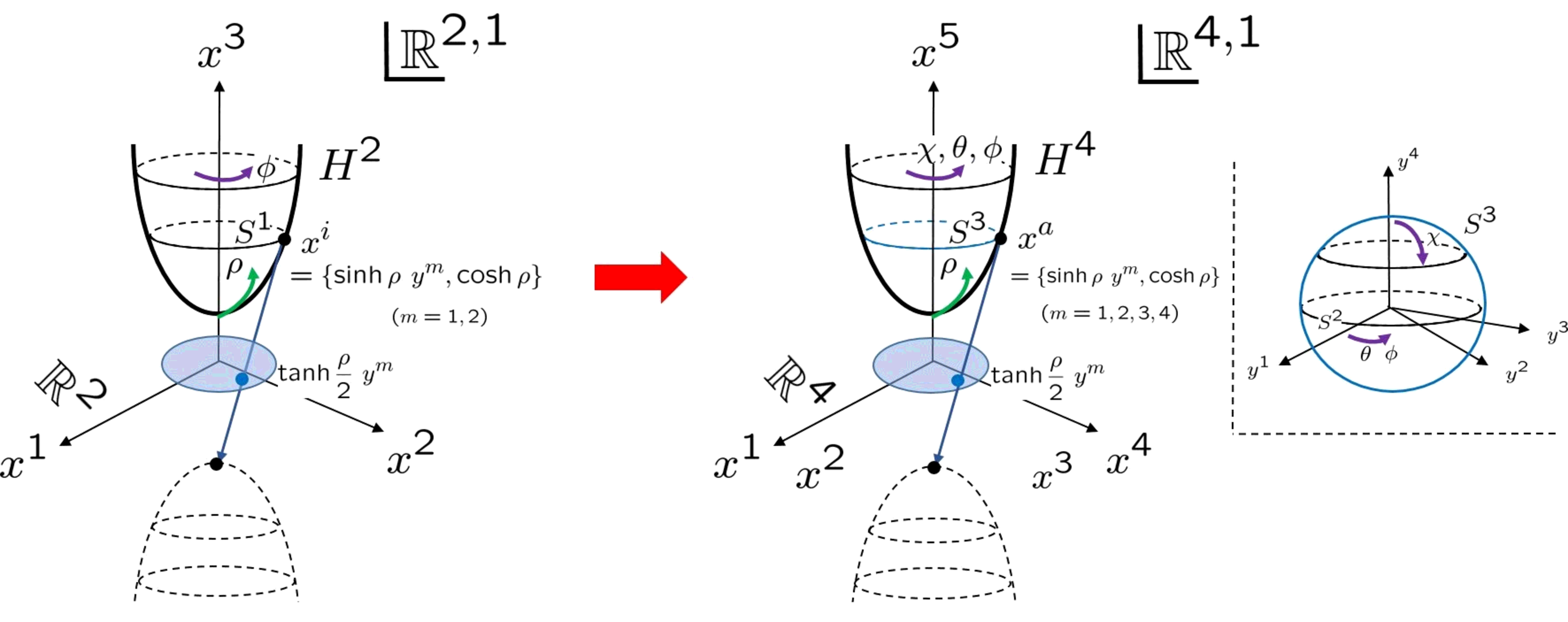}
\caption{2D hyperboloid $H^2$ (left),  4D hyperboloid $H^4$ (middle) and its $S^3$-latitude (right) }
\label{h4.fig}
\end{figure}

In this section,  we exploit the four-hyperboloid geometry (Fig.\ref{h4.fig}) and the $SO(4,1)$ coset representative. 
The geometric structures of  
$H^2$ and $H^4$ are respectively given by 
\be 
H^2~\simeq~\mathbb{R}^2
~\simeq~S^1 \otimes \mathbb{R}_+,~~~~~H^4~\simeq~\mathbb{R}^4
~\simeq~ S^3 \otimes \mathbb{R}_+. 
\ee 
$H^4$ can be   regarded as a natural generalization of $H^{2}$ whose latitude is expanded from $S^1$ to $S^3$. 
Such an $S^3$ structure  brings an internal spin structure to the $SO(4,1)$ formulation. 
Adopting a set of four Schwinger operators as an  $SO(4,1)$ representation, we derive an $SO(4,1)$ squeeze operator and discuss its internal spin structure.  

\subsection{$SO(4,1)$ algebra and Bloch four-hyperboloid}  

\subsubsection{$H^4$ and the $SO(4,1)$ algebra}

The four-hyperboloid $H^4$ is given by 
\be
\sum_{a,b=1}^5 \eta_{ab}x^a x^b =-\sum_{m=1}^4 x^mx^m +x^5 x^5 =1
\ee
with 
\be
\eta^{ab}=\eta_{ab}=\text{diag}(-,-,-,-,+). 
\ee
As a coset,  
$H^{4}$ is represented as 
\be
H^4 ~\simeq~SO(4,1)/SO(4).  
\ee
The $SO(4,1)$ gamma matrices $\gamma^a$ $(a=1,2,3,4,5)$ are introduced so as to satisfy 
\be
\{\gamma^a, \gamma^b\} =2\eta^{ab} ,  \label{so41gammaalgeb}
\ee
and the $SO(4,1)$ generators $\Sigma^{ab}$ are constructed as 
\be
\Sigma^{ab}=-i\frac{1}{4}[\gamma^a,\gamma^b],  
\ee 
which satisfy 
\be
[\Sigma^{ab}, \Sigma^{cd}] =i\eta^{ac}\Sigma^{bd}- i\eta^{ad}\Sigma^{bc}+ i\eta^{bd}\Sigma^{ac}- i\eta^{bc}\Sigma^{ad}. \label{so41genealge}
\ee
Hereafter, we take the $SO(4,1)$ gamma matrices and the generators  as 
\begin{align}
\gamma^{m}=
\begin{pmatrix}
0 & -\bar{q}^{m}\\
q^{m} & 0
\end{pmatrix},~~
\gamma^5=
\begin{pmatrix}
1_2 & 0  \\
0 & -1_2 
\end{pmatrix},  \label{so41gammamateken}
\end{align}
with  $q^m$ being quaternions, 
\be
q^m =\{-i\sigma_i, 1_2\}, ~~~\bar{q}^m =\{i\sigma_i, 1_2\}, 
\ee
and then\footnote{The independent generators of $\Sigma^{ab}$  are 
\be
-\frac{1}{2}
\begin{pmatrix}
\sigma_i & 0 \\
0 & 0
\end{pmatrix}, ~~~-\frac{1}{2}
\begin{pmatrix}
0 & 0 \\
0 & \sigma_i
\end{pmatrix},~~~-i\frac{1}{2}\begin{pmatrix}
0 & \bar{q}^{m}\\
q^{m} & 0 
\end{pmatrix}. \label{indo41gen}
\ee
Among $\Sigma^{ab}$ (\ref{so41matexpteke}), $\Sigma^{mn}$ are Hermitian matrices, while $\Sigma^{m 5}$ are anti-Hermitian matrices. $i\Sigma^{ab}$ (\ref{so41matexpteke}) are  recognized as the $U(1,1; \mathbb{H})$  generators, since their components are quaternions  
\be
i\Sigma^{ab}=\{\frac{1}{2}
\begin{pmatrix}
q_i & 0 \\
0 & 0
\end{pmatrix}, ~~~\frac{1}{2}
\begin{pmatrix}
0 & 0 \\
0 & q_i
\end{pmatrix},~~~\frac{1}{2}\begin{pmatrix}
0 & \bar{q}^{m}\\
q^{m} & 0 
\end{pmatrix}\},  \label{indo41genqs}
\ee
and they satisfy 
\be
(i\Sigma^{ab})^{\dagger}
\begin{pmatrix}
1 & 0 \\
0 & -1 
\end{pmatrix} = -\begin{pmatrix}
1 & 0 \\
0 & -1 
\end{pmatrix}i\Sigma^{ab}.
\ee
 } 
\begin{align}
&\Sigma^{mn} = -\frac{1}{2}
\begin{pmatrix}
\eta_{mn}^i\sigma_i & 0 \\
0 & \bar{\eta}_{mn}^i\sigma_i
\end{pmatrix}, ~~~\Sigma^{m 5}=-i\frac{1}{2}\begin{pmatrix}
0 & \bar{q}^{m}\\
q^{m} & 0 
\end{pmatrix}.  \label{so41matexpteke}
\end{align}
Here, $\eta_{mn}^i$ and $\bar{\eta}_{mn}^i$ are the 't Hooft symbols: 
\be
\eta_{mn}^i=\epsilon^{mn i 4}+\delta^{m i}\delta^{n 4}-\delta^{m 4}\delta^{n i}, ~~~\bar{\eta}_{mn}^i=\epsilon^{mn i 4}-\delta^{m i}\delta^{n 4}+\delta^{m 4}\delta^{n i}, \label{tHoofttensors}
\ee 
with $\epsilon^{1234}=1$. 
Notice that $\gamma^a$ and $\Sigma^{ab}$ amount to fifteen  matrices that satisfy the $SO(4,2)\simeq SU(2,2)$ algebra.  
With the $SU(2,2)$ invariant matrix 
\be
k=\gamma^5, 
\ee
we can ``hermitianize'' the $SU(2,2)$ generators as 
\be
k^a \equiv k\gamma^a,~~~k^{ab}=k\Sigma^{ab},  \label{kakabdef}
\ee
or 
\be
k^{m}=\begin{pmatrix}
0 & -\bar{q}^{m} \\
-q^{m} & 0 
\end{pmatrix}=\biggl\{\begin{pmatrix}
0 & -i\sigma_i \\
i\sigma_i & 0 
\end{pmatrix}, ~\begin{pmatrix}
0 & -1_2 \\
-1_2 & 0 
\end{pmatrix}\biggr\}, ~~~~~k^5=\begin{pmatrix}
1_2 & 0 \\
0 & 1_2 
\end{pmatrix}. 
\ee
Also note that neither the Hermitian matrices $k^a$ satisfy (\ref{so41gammaalgeb}) 
nor  $k^{ab}$  satisfy the $SO(4,1)$  algebra (\ref{so41genealge}). 

\subsubsection{Bloch four-hyperboloid and $SO(4,1)$ coset representative}

The  2nd  non-compact (hybrid) Hopf map   is  \cite{Hasebe-2012}
\be
H^{4,3} ~~\overset{S^3}{\longrightarrow} ~~H^{4}, 
\ee
which is explicitly given by 
\be
\Psi ~~\rightarrow~~x^a=\Psi^{\dagger}k^a \Psi.  \label{exp2ndhyb}
\ee
Here $\Psi=(\Psi_1~\Psi_2~\Psi_3~\Psi_4)^t$ denotes a four-component spinor subject to 
\be
\Psi^{\dagger}~k~\Psi =|\Psi_1|^2+|\Psi_2|^2-|\Psi_3|^2-|\Psi_4|^2 =1 ~:~H^{4,3}, 
\ee
and 
$x^a$ (\ref{exp2ndhyb}) automatically satisfy
\be
\eta_{ab}x^ax^b=-\sum_{m=1}^4 x^mx^m+x^5x^5 =(\Psi^{\dagger}k\Psi)^2=1 ~:~H^{4}.
\ee
We can realize $\Psi$ as 
\be
\Psi =\frac{1}{\sqrt{2(1+x^5)}}\begin{pmatrix}
(1+x^5) ~1_2 \\
x^mq_m
\end{pmatrix}\psi \label{cartcodpsi}
\ee
with two-component spinor $\psi=(\psi_1~\psi_2)^t$ subject to  
\be
\psi^{\dagger}\psi=1~:~S^3. 
\ee
With  polar angle coordinate representation\footnote{The  ranges of the parameters are 
\be
0\le  \rho< \infty,~~  0~ \le~  \chi, ~\theta~ \le~ \pi, ~~0~\le~\phi ~< 2\pi. 
\ee
}  
\begin{align}
&x^1=\sinh\rho\sin\chi\sin\theta\cos\phi,~~x^2=\sinh\rho\sin\chi\sin\theta\sin\phi,~~x^3=\sinh\rho\sin\chi\cos\theta,\nn\\
&x^4=\sinh\rho\cos\chi,~~~~~~~~x^5=\cosh\rho, 
\end{align}
(\ref{cartcodpsi}) is represented as 
\be
\Psi 
=\begin{pmatrix}
\cosh (\frac{\rho}{2}) ~1_2 \\
\sinh(\frac{\rho}{2}) ~y^mq_m
\end{pmatrix}\psi .  
\ee
Here, 
$y^m$ $(\sum_{m=1}^4 y^my^m =1)$ denote coordinates on the normalized $S^3$-``circle'' at latitude $\rho$ on $H^4$ (see Fig.\ref{h4.fig}): 
\be
y^m=\frac{1}{\sinh\rho}x^m=\{\sin\chi\sin\theta\cos\phi,~\sin\chi\sin\theta\sin\phi,~\sin\chi\cos\theta,~\cos\chi\}, 
\ee
and so 
\be
\sum_{m=1}^4 y^m q_m=-\sum_{m=1}^4 y^m q^m=
\begin{pmatrix}
- (\cos\chi-i\sin\chi\cos\theta) & i\sin\chi\sin\theta ~e^{-i\phi}  \\
i \sin\chi\sin\theta ~e^{i\phi} & - (\cos\chi+i\sin\chi\cos\theta)
\end{pmatrix}= (\sum_{m=1}^4 y^m \bar{q}_m)^{\dagger}.  \label{ymqmanglepara}
\ee
Notice that $\Psi$ acts as the $SO(4,1)$ pseudo-spin-coherent state on $H^{4}$: 
\be
(x^a \gamma_a) ~\Psi =\Psi, 
\ee
which suggests 
\be
\Psi=M(\rho,\chi,\theta,\phi)\begin{pmatrix}
\psi \\
0   
\end{pmatrix} , 
\ee
where $M(\rho,\chi,\theta,\phi)$ denotes an $SO(4,1)$ rotation\footnote{
In the polar coordinate representation, (\ref{ximatrix}) is  expressed as  
\begin{align}
&M(\rho, \chi, \theta,\phi) =
\nn\\
&\!\!\!\!\!\!\!\!\!\!\!\!\!\!\!\!\!\!\!\!\!\!\!\begin{pmatrix}
\cosh (\frac{\rho}{2})   &           0             &   -\sinh (\frac{\rho}{2})(\cos\chi+i\sin\chi\cos\theta)   &  -i\sinh (\frac{\rho}{2})\sin\chi\sin\theta ~e^{-i\phi} \\
0           & \cosh (\frac{\rho}{2})  &    -i\sinh (\frac{\rho}{2})   \sin\chi\sin\theta ~e^{i\phi} &  -\sinh (\frac{\rho}{2})(\cos\chi-i\sin\chi\cos\theta) \\ 
 -\sinh (\frac{\rho}{2})(\cos\chi-i\sin\chi\cos\theta)     &  i\sinh (\frac{\rho}{2})   \sin\chi\sin\theta ~e^{-i\phi} & \cosh (\frac{\rho}{2}) & 0  \\
i\sinh (\frac{\rho}{2})   \sin\chi\sin\theta ~e^{i\phi} & - \sinh(\frac{\rho}{2})(\cos\chi+i\sin\chi\cos\theta) )   & 0 & \cosh (\frac{\rho}{2})  
\end{pmatrix}. \label{squeezematexp}
\end{align}
}  
\be
M(\rho, \chi, \theta,\phi) 
= \begin{pmatrix}
\cosh (\frac{\rho}{2}) ~1_2 & \sinh (\frac{\rho}{2})\sum_{m=1}^4y^m\bar{q}_{m}\\
\sinh (\frac{\rho}{2}) \sum_{m=1}^4y^m q_{m} & \cosh (\frac{\rho}{2}) ~1_2 
\end{pmatrix} =e^{i\rho \sum_{m=1}^4 y^{m}\Sigma_{m 5}} , 
\label{ximatrix}
\ee 
with 
\be
\Sigma_{m5}=-\Sigma^{m 5} =i\frac{1}{2}\begin{pmatrix}
0 & \bar{q}^{m}\\
q^{m} & 0 
\end{pmatrix}. 
\ee
The last expression of (\ref{ximatrix}) implies $M(\rho, \chi, \theta,\phi)$ is a coset representative of $SO(4,1)$ group, since $\Sigma^{m5}$ act as the broken generators associated with the symmetry breaking,  
$SO(4,1)~~\rightarrow~~SO(4).$ 
Generalizing the $SO(5)$ coset matrix for  $S^4$ \cite{Hasebe-2020-1}, the Euler decomposition of (\ref{ximatrix})  
  is given by 
\be
M(\rho, \chi, \theta,\phi)=H(\chi,\theta,\phi)^{\dagger} \cdot e^{-i\rho\Sigma^{45}} \cdot  H(\chi,\theta,\phi).  
\label{eulerdecompsp41}
\ee 
Here, $H(\chi,\theta,\phi)$ denotes an $SU(2)$ matrix of the form 
\be
H(\chi,\theta,\phi) =e^{i\chi\Sigma^{34}} \cdot e^{i\theta\Sigma^{13}} \cdot e^{-i\phi\Sigma^{12}} =
\begin{pmatrix}
H_L(\chi,\theta,\phi) & 0 \\
0 &  H_R(\chi,\theta,\phi)
\end{pmatrix}, \label{totalsu2matdiag}
\ee
with 
\be
H_L(\chi,\theta,\phi)=H_R(-\chi,\theta,\phi) =e^{-i\frac{\chi}{2}\sigma_3}\cdot e^{i\frac{\theta}{2}\sigma_2} \cdot e^{i\frac{\phi}{2}\sigma_3} =\begin{pmatrix}
\cos(\frac{\theta}{2}) ~e^{-i\frac{1}{2}(\chi-\phi)} & \sin(\frac{\theta}{2}) ~e^{-i\frac{1}{2}(\chi+\phi)} \\
-\sin(\frac{\theta}{2}) ~e^{i\frac{1}{2}(\chi+\phi)} & \cos(\frac{\theta}{2}) ~e^{i\frac{1}{2}(\chi-\phi)} 
\end{pmatrix}. 
\ee 
Obviously (\ref{eulerdecompsp41}) is a natural generalization of   
the $SU(1,1)$ case (\ref{euler-decomp-sp2r}) with replacement of the $U(1)$ factor $e^{-i\frac{\phi}{2}\tau^z}$ with the $SU(2)$ factor $H(\chi,\theta,\phi)$ 
(\ref{totalsu2matdiag}). 
The Gauss decomposition 
is also given by 
\begin{align}
&~~M(\rho, \chi, \theta,\phi)=\exp\biggl(\tanh (\frac{\rho}{2}) ~ \begin{pmatrix}
0 & y^{m}\bar{q}_{m} \\
0 & 0 
\end{pmatrix} \biggr)\cdot \exp\biggl( -\ln (\cosh (\frac{\rho}{2}) )~ \begin{pmatrix} 
1_2 & 0 \\
0 & -1_2
\end{pmatrix}   \biggr)\cdot \exp\biggl(\tanh (\frac{\rho}{2}) ~ \begin{pmatrix}
0 & 0 \\
y^{m}{q}_{m} & 0 
\end{pmatrix} \biggr)\nn\\
&=\exp\biggl(-\tanh (\frac{\rho}{2}) ~  y^{m}\cdot (\frac{1}{2}\gamma_{m}-i\Sigma_{m 5}) \biggr)\cdot \exp\biggl( -\ln (\cosh (\frac{\rho}{2})  )  ~\gamma_5 \biggr)\cdot \exp\biggl(\tanh (\frac{\rho}{2})   ~y^{m} \cdot (\frac{1}{2}\gamma_{m}+i\Sigma_{m 5})\biggr), 
\label{gaussdecompsp41}
\end{align}
which realizes a natural generalization of  (\ref{gauss-decomp-sp2r}), again. 

\subsection{$SO(4,1)$ squeeze operator}

\subsubsection{$SO(4,1)$ Schwinger operator realization and  $SO(4,2)$ operators}

 As the $SO(4,1)$ group does not accommodate the Majorana representation,\footnote{
The $SO(4,1)$ gamma matrices (\ref{so41gammamateken}) and generators (\ref{indo41gen}) satisfy
\be
(\gamma^{a})^{*}=C \gamma^{a}C^{-1}, ~~~~(\Sigma^{ab})^* =-C\Sigma^{ab} C^{-1},  
\ee
where $C$ denotes  the charge conjugation matrix  
\be
C=
\begin{pmatrix}
0 & 1 & 0 & 0 \\
-1 & 0 & 0 & 0 \\
0 & 0 & 0 & 1 \\
0 & 0 & -1 & 0 
\end{pmatrix}. \label{chargecon}
\ee
With $C$ (\ref{chargecon})  the Majorana condition is imposed as 
\be
\Psi_{\text{M}}^*=C\Psi_\text{M}, 
\ee
and so 
\be
(\Psi_{\text{M}}^*)^*=C^* C\Psi_\text{M} =-\Psi_\text{M}. \label{majoranacons}
\ee
Under the usual definition of the complex conjugation $(\Psi^*)^*=\Psi$,  (\ref{majoranacons}) implies $\Psi_\text{M}=0$, which means that the Majorana spinor does not exist in the $SO(4,1)$ group.}  
  two-mode $SO(4,1)$ squeeze operator does not exist unlike the $SO(2,3)$ case \cite{Hasebe-2019}. 
We then consider  a set of  four Schwinger  operators in correspondence with the Dirac representation:  
\be
\hat{\Psi} =\begin{pmatrix}
a \\
b \\
c^{\dagger} \\
d^{\dagger}
\end{pmatrix}  \label{so41diracopspi}
\ee
whose components satisfy 
\be
[\hat{\Psi}_{\alpha}, \hat{\Psi}_{\beta}^{\dagger}] =k_{\alpha\beta}, ~~~[\hat{\Psi}_{\alpha}, \hat{\Psi}_{\beta}]=0. \label{comrepsi}
\ee
Sandwiching (\ref{kakabdef}) with (\ref{so41diracopspi}), we introduce the following  $\it{Hermitian}$ operators:  
\be
\hat{X}^a =\hat{\Psi}^{\dagger} k^a \hat{\Psi}, 
~~~\hat{X}^{ab} =\hat{\Psi}^{\dagger} k^{ab} \hat{\Psi}.
\ee 
$\hat{X}^a$ are explicitly 
\begin{align}
&\hat{X}^1 =-i(a^{\dagger} d^{\dagger} +b^{\dagger}c^{\dagger} -ad-bc), ~~\hat{X}^2 =-(a^{\dagger} d^{\dagger} -b^{\dagger}c^{\dagger} +ad-bc), \nn\\
&\hat{X}^3 =-i(a^{\dagger} c^{\dagger} -b^{\dagger}d^{\dagger} -ac+bd), ~~\hat{X}^4 =-(a^{\dagger} c^{\dagger} +b^{\dagger}d^{\dagger} +ac+bd), \nn\\
&\hat{X}^5 =a^{\dagger}a+ b^{\dagger}b +c^{\dagger}c +d^{\dagger}d +2, 
\end{align}
$\hat{X}^{ab}$ consist of the $SO(4)$ particle conserving operators $\hat{X}^{mn}$
\begin{align}
&\hat{X}^{12} =-\frac{1}{2}(a^{\dagger}a -b^{\dagger}b -c^{\dagger}c +d^{\dagger}d), ~~\hat{X}^{13} =i\frac{1}{2}(-a^{\dagger}b+b^{\dagger}a -c^{\dagger}d +d^{\dagger}c), \nn\\
&\hat{X}^{14} =-\frac{1}{2}(a^{\dagger}b +b^{\dagger}a +c^{\dagger}d +d^{\dagger}c),  ~~\hat{X}^{23} =\frac{1}{2}(a^{\dagger}b +b^{\dagger}a -c^{\dagger}d -d^{\dagger}c), \nn\\
&\hat{X}^{24} =-i\frac{1}{2}(a^{\dagger}b-b^{\dagger}a-c^{\dagger}d+d^{\dagger}c),~~\hat{X}^{34} =-\frac{1}{2}(a^{\dagger}a-b^{\dagger}b+c^{\dagger}c-d^{\dagger}d), 
\end{align}
and  the remaining  particle number non-conserving operators $\hat{X}^{m 5}$
\begin{align}
&\hat{X}^{15} =\frac{1}{2}(a^{\dagger}d^{\dagger}+b^{\dagger}c^{\dagger} +ad+bc),~~~\hat{X}^{25} =-i\frac{1}{2}(a^{\dagger}d^{\dagger}-b^{\dagger}c^{\dagger} -ad +bc),  \nn\\
&\hat{X}^{35} =\frac{1}{2}(a^{\dagger}c^{\dagger}-b^{\dagger}d^{\dagger} +ac -bd), ~~~\hat{X}^{45} =-i\frac{1}{2}(a^{\dagger}c^{\dagger}+b^{\dagger}d^{\dagger}-ac-bd). 
\end{align}
Notice that the  Hermitian matrices $k^{ab}$ (\ref{kakabdef}) do $\it{not}$ satisfy the $SO(4,1)$ algebra, but the corresponding Hermitian operators $\hat{X}^{ab}$ do satisfy the $SO(4,1)$ algebra. 
$SO(4,1)$ invariant operators are constructed as 
\be 
\hat{X}^a\hat{X}_a=\frac{1}{2}\sum_{a<b}\hat{X}^{ab}\hat{X}_{ab}
=({\hat{\Psi}}^{\dagger}k\hat{\Psi} ) ( {\hat{\Psi}}^{\dagger} k\hat{\Psi} +4 ), 
\ee 
where 
\be
{\hat{\Psi}}^{\dagger}k \hat{\Psi}=a^{\dagger}a +b^{\dagger}b -cc^{\dagger}-dd^{\dagger} =\hat{n}_a+\hat{n}_b-\hat{n}_c-\hat{n}_d-2.    
\ee
Furthermore, $\hat{X}^{ab}$ and $\hat{X}^{a 6}=-\hat{X}^{ 6 a}\equiv -\frac{1}{2}\hat{X}^a$ constitute the $SU(2,2)\simeq SO(4,2)$ algebra in total: 
\be
[\hat{X}^{AB}, \hat{X}^{CD}] =i\eta^{AC}\hat{X}^{BD}- i\eta^{AD}\hat{X}^{BC}+ i\eta^{BD}\hat{X}^{AC}- i\eta^{BC}\hat{X}^{AD},  ~~~(A,B,C,D=1,2,3,4,5,6)
\ee
with  
\be
\eta^{AB}=\text{diag}(-,-,-,-,+,+). 
\ee
The corresponding $SU(2,2)~\simeq~SO(4,2)$ Casimir is constructed as  
\be
\hat{X}^a\hat{X}_a+4\sum_{a<b}\hat{X}^{ab}\hat{X}_{ab}= 3 (\hat{n}_a+\hat{n}_b-\hat{n}_c-\hat{n}_d-2)(\hat{n}_a+\hat{n}_b-\hat{n}_c-\hat{n}_d+2)=3({\hat{\Psi}}^{\dagger}k\hat{\Psi} )({\hat{\Psi}}^{\dagger}k\hat{\Psi} +4) . \label{su22casimir}
\ee 
The eigenstates are given by\footnote{
For oscillator realization of the $SU(2,2)$ group and its representation, one may consult Refs.\cite{Barut-Bohm-1970, Gunaydin-Saclioglu-1982-1,Gunaydin-Saclioglu-1982-2,Bars-Gunaydin-1983,Bars-Gunaydin-1998,Fernando-Gunaydin-2010,Govil-Gunaydin-2015}.   
}  
\be
|n_a, n_b, n_c, n_d\rangle =\frac{1}{\sqrt{n_a! n_b! n_c! n_d!}} ~(a^{\dagger})^{n_a}~(b^{\dagger})^{n_b}~(c^{\dagger})^{n_c}~(d^{\dagger})^{n_d}~ |0,0,0,0\rangle . \label{su22eigenstates}
\ee
Obviously, the particle numbers, $n_a$, $n_b$, $n_c$ and $n_d$, uniquely specify the eigenstates.

\subsubsection{Subgroup decomposition of the $SO(4,1)$ squeeze operator }

From a view of the $SO(4,2)$ group theory,   we delve into internal spin and pseudo-spin structures.  
The $SO(4,2)$ group  accommodates two important  subgroups, the $SO(4)$ group for spins  and the $SO(2,2)$ group for pseudo-spins:  
\begin{subequations}
\begin{align}
&~~SO(4)~~\simeq~~SU(2)_L\otimes SU(2)_R ~~\subset ~~~SO(4,1)~~\subset ~~SO(4,2),  \\
&SO(2,2)~\simeq~SU(1,1)_T\otimes SU(1,1)_B 
~\subset ~SO(2,3) ~\subset ~SO(4,2).  \label{so42toso22}
\end{align}
\end{subequations}
For $SO(4)~\simeq~SU(2)_L\otimes SU(2)_R$, the corresponding left and right $SU(2)$ spin operators are constructed as 
\be
\hat{L}_i =\frac{1}{4}\eta^i_{mn} \hat{X}^{mn} =\begin{pmatrix}
a \\
b 
\end{pmatrix}^{\dagger} \frac{1}{2}\sigma_i  \begin{pmatrix}
a \\
b 
\end{pmatrix}, ~~~
\hat{R}_i =\frac{1}{4}\bar{\eta}^i_{mn} \hat{X}^{mn} =\begin{pmatrix}
c \\
d 
\end{pmatrix}^{\dagger} \frac{1}{2}\sigma_i  \begin{pmatrix}
c \\
d 
\end{pmatrix}, \label{deflandrsops}
\ee 
which satisfy  
\be
\sum_{i=1}^3 {\hat{L}_i}{\hat{L}_i}=\hat{L}(\hat{L}+1), ~~~\sum_{i=1}^3 {\hat{R}_i}{\hat{R}_i}=\hat{R}(\hat{R}+1) 
\ee
with 
\be
\hat{L}=\frac{1}{2}(\hat{n}_a+\hat{n}_b),~~\hat{R}=\frac{1}{2}(\hat{n}_c+\hat{n}_d). 
\ee
The $SU(2,2)$ eigenstate (\ref{su22eigenstates}) can be  specified by 
the spin group quantum numbers, 
\be
\hat{L}=\frac{1}{2}(\hat{n}_a+\hat{n}_b), ~~~\hat{R}=\frac{1}{2}(\hat{n}_c+\hat{n}_d), ~~\hat{L}_z=\frac{1}{2}(\hat{n}_a-\hat{n}_b), ~~~\hat{R}_z=\frac{1}{2}(\hat{n}_c-\hat{n}_d). 
\ee 
Meanwhile for $SO(2,2)~\simeq~SU(1,1)_T\otimes SU(1,1)_B$, the corresponding top and bottom $SU(1,1)$ pseudo-spin operators are introduced as 
\be
\hat{T}^i =\begin{pmatrix}
a \\
c^{\dagger}
\end{pmatrix}^{\dagger} \frac{1}{2}\kappa^i \begin{pmatrix}
a \\
c^{\dagger}
\end{pmatrix}, ~~~\hat{B}^i =\begin{pmatrix}
b \\
d^{\dagger}
\end{pmatrix}^{\dagger} \frac{1}{2}\kappa^i \begin{pmatrix}
b \\
d^{\dagger}
\end{pmatrix}, 
\ee
which satisfy  
\be
\sum_{i,j=1}^3 \eta_{ij}{\hat{T}^i}{\hat{T}^j}=\hat{T}(\hat{T}+1), ~~~\sum_{i=1}^3 \eta_{ij} {\hat{B}^i}{\hat{B}^j}=\hat{B}(\hat{B}+1), 
\ee
with $\eta_{ij}=\text{diag}(-,-,+)$  and 
\be
\hat{T} =\frac{1}{2}(\hat{n}_a-\hat{n}_c-1),~~~\hat{B}=\frac{1}{2}(\hat{n}_b-\hat{n}_d-1). 
\ee
In detail, 
\begin{align}
&\hat{T}^x =i\frac{1}{2}(a^{\dagger}c^{\dagger}-ac) =\frac{1}{2}X^{45} -\frac{1}{4}X^3, ~~\hat{T}^y =\frac{1}{2}(a^{\dagger}c^{\dagger}+ac) =-\frac{1}{2}\hat{X}^{35} -\frac{1}{4}\hat{X}^4, \nn\\
&\hat{T}^z =\frac{1}{2}(a^{\dagger}a +c^{\dagger}c)+\frac{1}{2} =\frac{1}{2}\hat{X}^{34} +\frac{1}{4}\hat{X}^5, ~~\hat{B}^x =i\frac{1}{2}(b^{\dagger}d^{\dagger}-bd) =\frac{1}{2}X^{45} +\frac{1}{4}X^3, \nn\\
&\hat{B}^y =\frac{1}{2}(b^{\dagger}d^{\dagger}+bd) =\frac{1}{2}\hat{X}^{35} -\frac{1}{4}\hat{X}^4, ~~\hat{B}^z =\frac{1}{2}(b^{\dagger}b +d^{\dagger}d)+\frac{1}{2} =-\frac{1}{2}\hat{X}^{34} +\frac{1}{4}\hat{X}^5. 
\end{align}
The $SU(2,2)$ eigenstate (\ref{su22eigenstates}) can be  specified by the pseudo-spin quantum numbers of $\hat{T}$, $\hat{B}$, $\hat{T}^z$ and $\hat{B}^z$.

\subsubsection{$SO(4,1)$  squeeze operator and the interaction Hamiltonian}\label{subsec:so41sqopham}

The existence of  spin and pseudo-spin groups in the $SO(4,1)$ group implies   that the $SO(4,1)$ formalism is large enough to incorporate  both spin  and   squeezed state structures.   
Replacing the non-Hermitian matrices in  (\ref{ximatrix}) with the corresponding Hermitian operators, we introduce the $SO(4,1)$ squeeze operator: 
\be
\hat{S}(\rho,\chi,\theta,\phi) =e^{i\rho \sum_{m=1}^4 y^{m}(\chi,\theta,\phi) \hat{X}_{m 5}}=e^{-i\rho \sum_{m=1}^4 y^{m}(\chi,\theta,\phi) \hat{X}^{m 5}}. 
\label{so41squop}
\ee
The interaction Hamiltonian 
is then given by  
\begin{align}
&\hat{H}^{\text{int}}(\rho, \chi, \theta, \phi) =- \rho\sum_{m=1}^4 y^m \hat{X}_{m5} 
\nn\\
&~~~~~~~~~=\frac{\rho}{2}\biggl(\sin\chi~(\cos\theta (a^{\dagger}c^{\dagger}-b^{\dagger}d^{\dagger})+\sin\theta (e^{-i\phi} a^{\dagger}d^{\dagger} +e^{i\phi} b^{\dagger}c^{\dagger}))-i\cos\chi~ (a^{\dagger}c^{\dagger}+b^{\dagger}d^{\dagger}) \biggr) + \text{h.c.}  \label{so41hamintcoord}
\end{align}
The $SO(4,1)$ squeeze operator is a unitary operator that satisfies 
\be
\hat{S}(\rho, \chi, \theta, \phi)^{\dagger} = \hat{S}(\rho, \chi, \theta, \phi)^{-1}  = \hat{S}(-\rho, \chi, \theta, \phi).
\ee
In (\ref{so41hamintcoord}), we can easily see that the phase $\phi$ can be absorbed in the phase redefinition of the Schwinger operators: 
\be
a ~~\rightarrow~~e^{-i\alpha}~a,~~~b ~~\rightarrow~~e^{i\beta}~b, ~~~~c ~~\rightarrow~~e^{i\alpha}~c,~~~d ~~\rightarrow~~e^{-i\beta}~d,\label{redaadaggerbcd}
\ee
where 
\be
\alpha+\beta=\phi. \label{sumanglesphi}
\ee
To grasp physical meaning of the $SO(4,1)$ squeeze operator (\ref{so41squop}), we rewrite the $SO(4,1)$ squeeze operator in the language of the $SU(2)$ spin and the $SU(1,1)$ pseudo-spin groups. For this purpose, we apply the  Euler decomposition (\ref{eulerdecompsp41}) to the squeeze operator  (\ref{so41squop}):
\footnote{
One may of course utilize the Gauss decomposition of $\hat{S}(\rho,\chi,\theta,\phi) $ (\ref{gaussdecompsp41}) in deriving the number state expansion of the  $SO(4,1)$ squeezed vacuum: 
\be
\hat{S}(\rho,\chi,\theta,\phi)  =\exp\biggl(-\tanh(\frac{\rho}{2})~y^m (\frac{1}{2}\hat{X}_m-i\hat{X}_{m5})\biggr)\cdot \exp\biggl(-\log (\cosh(\frac{\rho}{2}) ~\hat{X}_5 )   \biggr) \cdot \exp\biggl(\tanh(\frac{\rho}{2})~y^m (\frac{1}{2}\hat{X}_m+i\hat{X}_{m5})\biggr), 
\ee
where 
\begin{align}
&y^m (\frac{1}{2}\hat{X}_m-i\hat{X}_{m5}) =i\sin\chi\sin\theta e^{-i\phi}a^{\dagger}d^{\dagger}+i\sin\chi\sin\theta e^{i\phi}b^{\dagger}c^{\dagger}+(\cos\chi+i\sin\chi\cos\theta)a^{\dagger}c^{\dagger} + (\cos\chi-i\sin\chi\cos\theta)b^{\dagger}d^{\dagger} , \nn\\
&\hat{X}_5=\hat{n}_a+\hat{n_b}+\hat{n}_c+\hat{n}_d+2.   
\end{align}
In the main text, we however adopted the Euler decomposition as it manifests a clear physical meaning of the $SO(4,1)$ squeezed vacuum. 
} 
\be
\hat{S}(\rho,\chi,\theta,\phi)= \hat{H}(\chi,\theta,\phi)^{\dagger}\cdot \hat{S}(\rho) \cdot \hat{H}(\chi,\theta,\phi), 
\label{diractypesq}
\ee 
where 
\be
 \hat{H}(\chi,\theta,\phi)\equiv e^{i\chi \hat{X}^{34}} \cdot e^{i\theta \hat{X}^{13}} \cdot e^{-i\phi \hat{X}^{12}} =\hat{D}_L(\chi,-\theta,-\phi)\cdot \hat{D}_R(\chi,-\theta,\phi) 
\ee
and 
\be
\hat{S}(\rho)\equiv  e^{-i\rho \hat{X}^{45}} =\hat{S}_T(\rho )  \cdot \hat{S}_B(\rho ).  
\ee
Here, 
\be
\hat{D}_L(\chi,\theta,\phi) \equiv e^{-i\chi\hat{L}^z} e^{-i\theta\hat{L}^y} e^{-i\phi\hat{L}^z} , ~~\hat{D}_R(\chi,\theta,\phi) \equiv e^{-i\chi\hat{R}^z} e^{-i\theta\hat{R}^y} e^{-i\phi\hat{R}^z},  ~~\hat{S}_T(\rho ) \equiv e^{i\rho \hat{T}^x}, ~~\hat{S}_B(\rho ) \equiv e^{i\rho \hat{B}^x}. 
\ee 
Consequently, the $SO(4,1)$ squeeze operator can be expressed as\footnote{  
We used 
$\hat{D}(\chi, \theta, \phi)^{\dagger} = \hat{D}(-\phi, -\theta, -\chi)$ in the derivation of  (\ref{so41sqdecomp}).  
} 
\be
\hat{S}(\rho,\chi,\theta,\phi)  =  \hat{D}_L(\phi,\theta,-\chi)\cdot \hat{D}_R(-\phi,\theta,-\chi)  \cdot  \hat{S}_T(\rho )\cdot \hat{S}_B(\rho )\cdot \hat{D}_L(\chi,-\theta,-\phi)\cdot \hat{D}_R(\chi,-\theta,\phi), \label{so41sqdecomp}
\ee
Removing the right $SU(2)$ factor of (\ref{diractypesq}), we introduce the Schwinger-type $SO(4,1)$ squeeze operator:\footnote{ The corresponding squeeze matrix of the Schwinger-type is given by  
\begin{align}
\mathcal{M}(\rho,\chi,\theta,\phi) &=H(\chi,\theta,\phi)^{\dagger} \cdot e^{-i\rho\Sigma^{45}}  
\nn\\
& \!\!\!\!\!\!\!\!\!\!\!\!\!\! =\begin{pmatrix}
\cosh \frac{\rho}{2} ~\cos\frac{\theta}{2} ~ e^{i\frac{1}{2}(\chi-\phi)} &  -\cosh \frac{\rho}{2} ~\sin\frac{\theta}{2} ~ e^{-i\frac{1}{2}(\chi+\phi)} &  -\sinh \frac{\rho}{2} ~\cos\frac{\theta}{2} ~ e^{i\frac{1}{2}(\chi-\phi)} &  \sinh \frac{\rho}{2} ~\sin\frac{\theta}{2} ~ e^{-i\frac{1}{2}(\chi+\phi)} \\ 
\cosh \frac{\rho}{2} ~\sin\frac{\theta}{2} ~ e^{i\frac{1}{2}(\chi+\phi)} &  \cosh \frac{\rho}{2} ~\cos\frac{\theta}{2} ~ e^{-i\frac{1}{2}(\chi-\phi)} &  -\sinh \frac{\rho}{2} ~\sin\frac{\theta}{2} ~ e^{i\frac{1}{2}(\chi+\phi)} &  -\sinh \frac{\rho}{2} ~\cos\frac{\theta}{2} ~ e^{-i\frac{1}{2}(\chi-\phi)} \\ 
-\sinh \frac{\rho}{2} ~\cos\frac{\theta}{2} ~ e^{-i\frac{1}{2}(\chi+\phi)} &  \sinh \frac{\rho}{2} ~\sin\frac{\theta}{2} ~ e^{i\frac{1}{2}(\chi-\phi)} &  \cosh \frac{\rho}{2} ~\cos\frac{\theta}{2} ~ e^{-i\frac{1}{2}(\chi+\phi)} &  -\cosh \frac{\rho}{2} ~\sin\frac{\theta}{2} ~ e^{i\frac{1}{2}(\chi-\phi)} \\ 
-\sinh \frac{\rho}{2} ~\sin\frac{\theta}{2} ~ e^{-i\frac{1}{2}(\chi-\phi)} &  -\sinh \frac{\rho}{2} ~\cos\frac{\theta}{2} ~ e^{i\frac{1}{2}(\chi+\phi)} &  \cosh \frac{\rho}{2} ~\sin\frac{\theta}{2} ~ e^{-i\frac{1}{2}(\chi-\phi)} &  \cosh \frac{\rho}{2} ~\cos\frac{\theta}{2} ~ e^{i\frac{1}{2}(\chi+\phi)} 
\end{pmatrix}. \label{schwingertypema} 
\end{align}
}    
\be
\hat{\mathcal{S}}(\rho,\chi,\theta,\phi)  =\hat{H}(\chi,\theta,\phi)^{\dagger}\cdot\hat{S}(\rho)=\hat{D}_L(\phi,\theta,-\chi)\cdot \hat{D}_R(-\phi,\theta,-\chi)  \cdot  \hat{S}_T(\rho )\cdot \hat{S}_B(\rho ). 
\label{schwingertypesq}
\ee 

\section{$SO(4,1)$ squeezed vacuum and its basic properties}\label{sec:so41sqvac}

Armed with the spin and pseudo-spin based expression of the $SO(4,1)$ squeeze operator, we now derive a mathematical expression of the $SO(4,1)$ squeezed vacuum and discuss  its physical properties.   

\subsection{$SO(4,1)$ squeezed vacuum}

\subsubsection{Spin interpretation}

\begin{figure}[tbph]
\center
\hspace{-1.0cm}
\includegraphics*[width=140mm]{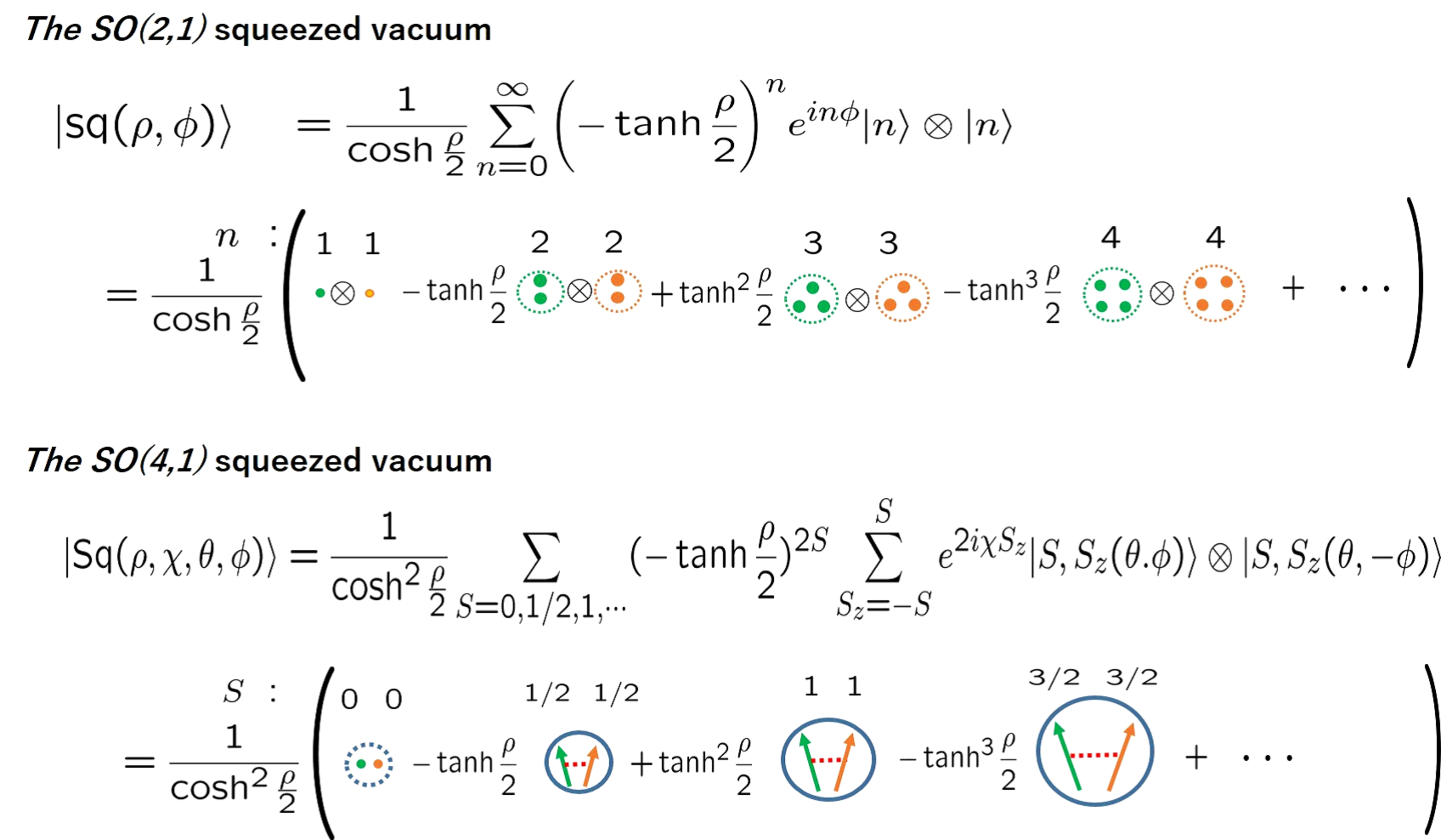}
\caption{The original $SO(2,1)$ squeezed vacuum is an entangled state of infinite (separable) pairs of number-states. 
The $SO(4,1)$ squeezed vacuum is an entangled state of  infinite   (maximally entangled) spin-pairs. 
}
\label{spinexpan.fig}
\end{figure}

We simply construct the $SO(4,1)$ squeezed vacuum by applying the $SO(4,1)$ squeeze operator to the vacuum: 
\be
|\text{Sq}(\rho,\chi,\theta,\phi)\rangle =\hat{S}(\rho,\chi,\theta,\phi) |0,0,0,0\rangle. \label{simpledefsquso41}
\ee
The squeezed vacuum is identical for the Dirac-type $\hat{S}$ and the Schwinger-type $\hat{\mathcal{S}}$:
\begin{align}
|\text{Sq}(\rho,\chi,\theta,\phi)\rangle 
&=\hat{D}_L(\phi,\theta, -\chi) \cdot \hat{D}_R (-\phi,\theta, -\chi) \cdot \hat{S}_T(\rho)\cdot \hat{S}_B(\rho)|0,0,0,0\rangle \nn\\
&=\hat{\mathcal{S}}(\rho,\chi,\theta,\phi)|0,0,0,0\rangle.  
\label{sqvacorig}
\end{align}
This is a property similar to  the original $SO(2,1)$ case,  
 in which the squeezed vacua in the Dirac-type and the Schwinger-type are physically equivalent (\ref{sqsqdiff}). 
  (In the case of the $Sp(4; \mathbb{R})\simeq SO(2,3)$ squeezed states \cite{Hasebe-2019},   the Dirac-type  and Schwinger-type squeezed vacua are physically different.)  
Thus, the    $SO(4,1)$ squeezed vacuum inherits the uniqueness  of the squeezed vacuum.     
Expanding the two squeeze operators, $\hat{S}_T$ and $\hat{S}_B$, in terms of the number operators,  we can express the squeezed vacuum (\ref{sqvacorig}) as 
\be 
|\text{Sq}(\rho, \chi, \theta, \phi)\rangle 
= \frac{1}{\cosh^2(\frac{\rho}{2})}\sum_{n,m=0}^{\infty} \biggl(-\tanh (\frac{\rho}{2})\biggr)^{n+m}~\hat{D}_L(\phi,\theta, -\chi) \hat{D}_R (-\phi,\theta, -\chi)~|n,m\rangle  \otimes|n,m\rangle. \label{squvac}
\ee
With the spin interpretation (\ref{angunnequiv}), we can rewrite (\ref{squvac})   as 
\begin{align}
|\text{Sq}(\rho, \chi, \theta, \phi)\rangle &= \frac{1}{\cosh^2(\frac{\rho}{2})} \sum_{S=0,1/2,1,3/2, \cdots}\sum_{S_z=-S}^{S} \biggl(-\tanh (\frac{\rho}{2})\biggr)^{2S}~\hat{D}_L(\phi,\theta, -\chi) \cdot \hat{D}_R (-\phi,\theta, -\chi)~|S,S_z\rangle \otimes |S,S_z\rangle  \nn\\
&= \frac{1}{\cosh^2(\frac{\rho}{2})}\sum_{S=0,1/2,1,3/2, \cdots} \biggl(-\tanh (\frac{\rho}{2})\biggr)^{2S}~\sum_{S_z=-S}^{S} ~|S,S_z (\phi,\theta, -\chi) \rangle \otimes |S,S_z (-\phi,\theta, -\chi) \rangle  .  \label{spinexpsequso41p}
\end{align}
Here, we changed the summation from $n$ and $m$ to $S$ and $S_z$ according to (\ref{changenmtossz}). 
Thus, 
the $SO(4,1)$ $\it{four}$-mode squeezed vacuum  can be understood   as a linear combination of  $\it{two}$-body spin-coherent states. 
In (\ref{spinexpsequso41p})  the first spin direction is specified by $(\theta, -\phi)$ while the second spin direction is specified by $(\theta, \phi)$, and then, by tuning $\theta$ and $\phi$,  we can manipulate relative configurations of two spins.\footnote{In the four-mode interpretation, due to the independent phase choice of the four-mode operators, the angle $\phi$  can be  ``gauged''  away (\ref{sumanglesphi}), but in the spin interpretation, only the  overall phase of $a$ and $b$ and that of $c$ and $d$ are allowed to change, since the angular momentum operators (\ref{deflandrsops}) are just immune to overall phase redefinition of these operators.  Thus in the spin interpretation, $\phi$ cannot be gauged away and  is indeed related the relative direction of two spins.  }  
Extracting the $U(1)$ phase factor of the spin state 
\be
|S, S_z (\phi,\theta, \chi)\rangle =e^{-i\chi S_z}|S, S_z(\theta, \phi)\rangle, 
\ee
we can express (\ref{spinexpsequso41p}) as 
\be 
|\text{Sq}(\rho, \chi, \theta, \phi)\rangle= \frac{1}{\cosh^2(\frac{\rho}{2})}\sum_{S=0,1/2,1,3/2, \cdots} \biggl(-\tanh (\frac{\rho}{2})\biggr)^{2S}~\sum_{S_z=-S}^{S} ~e^{2i\chi S_z}|S,S_z (\theta, \phi) \rangle \otimes |S,S_z (\theta, -\phi) \rangle,  \label{spinexpsequso41}
\ee 
where 
\be
|S, S_z (\theta, \phi)\rangle \equiv  \hat{D}(\phi,\theta, 0)|S,S\rangle. 
\ee
 One may find that 
(\ref{spinexpsequso41}) 
 naturally generalizes  the number-state expansion of the  original squeezed vacuum (\ref{numberstaterepsq2r}) [Fig.\ref{spinexpan.fig}].  

\subsubsection{Spin entanglement}

We can represent 
(\ref{spinexpsequso41p}) as  
\be 
|\text{Sq}~(\rho, \chi, \theta, \phi)\rangle =\frac{1}{\cosh^2(\frac{\rho}{2})}~\sum_{S=0, 1/2, 1, 3/2, \cdots }\sqrt{2S+1}~\biggl(-\tanh (\frac{\rho}{2})\biggr)^{2S} ~|S\otimes S ~ (\chi, \theta, \phi)\rangle\!\rangle,  \label{simpleexpnumbs}
\ee 
where $|S\otimes S ~ (\chi, \theta, \phi)\rangle$ denotes a (normalized) spin-pair state:  
\begin{align} 
|S\otimes S ~ (\chi, \theta, \phi)\rangle\!\rangle &\equiv \frac{1}{\sqrt{2S+1}}~\sum_{S_z=-S}^S~|S,S_z (\phi,\theta, -\chi) \rangle \otimes |S,S_z (-\phi,\theta, -\chi) \rangle \nn\\ 
&= \frac{1}{\sqrt{2S+1}} ~\sum_{S_z=-S}^S\sum_{L_z=-S}^S \sum_{R_z=-S}^S D^{(S)}(\phi,\theta,-\chi)_{L_z,S_z} D^{(S)}(-\phi,\theta,\chi)_{R_z,S_z}  ~|S, L_z\rangle \otimes |S, R_z  \rangle \nn \\
&=\frac{1}{\sqrt{2S+1}}~\sum_{L_z=-S}^S \sum_{R_z=-S}^S ~(D^{(S)}(\phi, \theta, -\chi) ~D^{(S)}(-\chi, -\theta, -\phi))_{L_z, R_z} ~|S, L_z\rangle \otimes |S, R_z  \rangle  .  \label{stimescjhirhophi}
\end{align}
In the last equation, we used the property of the D-matrix: 
\be
D^{(S)} (\chi,\theta, \phi)^t =D^{(S)}(\phi, -\theta, \chi). 
\ee
(\ref{simpleexpnumbs}) exhibits a duplicate entanglement which we refer to as the hybrid entanglement:  
\begin{enumerate}
\item  An infinite number of spin-pairs  are entangled with the squeeze parameter $\rho$. This manifests a natural 
  generalization of the entanglement of the original $SO(2,1)$ squeezed vacuum. 
\item In each spin-sector, two spins  are maximally entangled. This is a unique feature in the $SO(4,1)$ case. In the original $SO(2,1)$ squeezed vacuum, the corresponding state  is simply a product state, $|n\rangle\otimes |n\rangle$.  
\end{enumerate}
The spin part (\ref{stimescjhirhophi}) can be expressed as a superposition of the  integer total spin states only:  
\be
I\equiv S\otimes S =0 \oplus 1 \oplus  2 \oplus  \cdots \oplus  2S. 
\ee
This is a generalization of the fact that  even total number  states $|n\rangle \otimes |n\rangle$ 
only appear in the expansion of the two-mode $SO(2,1)$ squeezed vacuum.  
Using the Clebsch-Gordan coefficient $C^{j, j_z}_{j_1,m_1; j_2, m_2}$, we can construct total integer spin states  
\be
|I, I_z\rangle \!\rangle =  \sum_{L_z=-S}^S   \sum_{R_z=-S}^S C_{S, L_z; S, R_z}^{I, I_z} |S,L_z\rangle \otimes |S, R_z\rangle, 
\ee
and  (\ref{stimescjhirhophi}) is expressed as 
\be
|S\otimes S~ (\chi, \theta, \phi)\rangle\!\rangle 
=\frac{1}{\sqrt{2S+1}}  \sum_{I=0}^{2S} \sum_{I_z=-I}^I   C_{S}^{I, I_z} (\chi,\theta,\phi)|I, I_z\rangle \!\rangle
\ee 
where 
\be
C_{S}^{I, I_z} (\chi,\theta,\phi)\equiv \sum_{L_z=-S}^S \sum_{R_z=-S}^S  C_{S, L_z; S, R_z}^{I, I_z} (D^{(S)}(\phi, \theta, -\chi) ~D^{(S)}(-\chi, -\theta, -\phi))_{L_z, R_z}.  
\ee
Consequently, the $SO(4,1)$ squeezed vacuum (\ref{simpleexpnumbs}) is actually expressed as  a linear combination of an infinite number of spin-pairs with  total  integer spins: 
\be
|\text{Sq}~(\rho, \chi, \theta, \phi)\rangle 
= \sum_{I=0, 1, 2, \cdots }~ \sum_{I_z=-I}^I   ~C^{I, I_z} (\rho, \chi,\theta,\phi)~|I, I_z\rangle \!\rangle  , 
\label{ortan1sq}
\ee
with  
\be
C^{I, I_z} (\rho, \chi,\theta,\phi)\equiv ~\frac{1}{\cosh^2(\frac{\rho}{2})}~\sum_{S=\frac{I}{2}, \frac{I}{2}+1, \frac{I}{2}+2, \cdots }~(-\tanh (\frac{\rho}{2}))^{2S}~ C_{S}^{I, I_z} (\chi,\theta,\phi). 
\ee
Up to $O(\tanh(\frac{\rho}{2}))$,   (\ref{ortan1sq}) is  represented as   
\begin{align}
|\text{Sq}~(\rho,\chi,\theta,\phi)\rangle  &
=\frac{1}{\cosh^2(\frac{\rho}{2})} \biggl( |0,0\rangle \!\rangle -\tanh(\frac{\rho}{2})    \biggl((\cos\chi+i\sin\chi\cos\theta)|1,1\rangle \!\rangle +\sqrt{2} i\sin\chi\sin\theta\cos\phi |1,0\rangle \!\rangle \nn\\
&~~~~~~~~~~~~+(\cos\chi-i\sin\chi\cos\theta)|1,-1\rangle \!\rangle  +\sqrt{2}\sin\chi\sin\theta\sin\phi|0,0\rangle\!\rangle \biggr) +O(\tanh^2(\frac{\rho}{2})) \biggr).
\end{align}

\subsection{Dimensional Reduction}

Here, we discuss   reductions of the squeezed vacuum associated with lower dimensional geometries of the Bloch four-hyperboloid.  
First, we consider a 0D reduction: We focus on the the lowest point of the upper leaf of $H^4$, $i.e.$, $(x^1, x^2, x^3 , x^4, x^5)=(0,0,0,0,1)$, which corresponds to no squeezing $\rho=0$, and so the squeeze operator (\ref{so41squop}) becomes trivial: 
\be
\hat{S} =1. 
\ee
The squeezed vacuum (\ref{spinexpsequso41}) is reduced to the trivial vacuum: 
\be
|\text{Sq} \rangle 
= |0,0,0,0\rangle. 
\ee
Next, let us consider a 2D reduction,  $H^{4} ~~\rightarrow ~~H^2$, which is realized at $\theta=0$:  
\be 
(x^1, x^2, x^3 , x^4, x^5)=(0,0,\sin\chi\sinh\rho,\cos\chi\sinh\rho,\cosh\rho).  \label{h2coordexpcar}
\ee
The interaction Hamiltonian (\ref{so41hamintcoord}) is reduced to 
\begin{align}
\hat{H}^{\text{int}}(\rho, \chi, 0, \phi) &=-i\frac{\rho}{2}a^{\dagger}c^{\dagger}~e^{i\chi}-i\frac{\rho}{2}b^{\dagger}d^{\dagger}~e^{-i\chi} +\text{h.c.} =\hat{H}_{T}^{\text{int}}(\rho,\chi) +\hat{H}_{B}^{\text{int}}(\rho, -\chi)\nn\\
&=\hat{H}^{\text{int}}(\rho, \chi, 0, 0),  \label{rhochiintham}
\end{align}
and then the squeezed vacuum (\ref{spinexpsequso41}) becomes   
\be
|\text{Sq} \rangle 
= |\text{sq}(\rho, \chi)\rangle_{T}\otimes |\text{sq}(\rho, -\chi)\rangle_{B} 
=\frac{1}{\cosh^2(\frac{\rho}{2})}\sum_{S=0, 1/2, 1, \cdots} \biggl(-\tanh(\frac{\rho}{2})\biggr)^{2S}\sum_{S_z=-S}^S e^{2i\chi S_z}|S,S_z\rangle \otimes |S, S_z\rangle, \label{tbsqueezedstates}
\ee
which is obvious from  (\ref{prodtwomodesqussss}). 
Lastly, let us consider 1D reductions. 
For 
$\chi=0$, we have  $(x^1,x^2,x^3, x^4, x^5)=(0,0,0, \sinh\rho, \cosh\rho)$  
and the squeezed state is reduced to 
\be 
|\text{Sq} \rangle=\frac{1}{\cosh^2(\frac{\rho}{2})}\sum_{S=0, 1/2, 1, \cdots }\sqrt{2S+1}(-\tanh(\frac{\rho}{2}))^{2S}\cdot \frac{1}{\sqrt{2S+1}}\sum_{S_z=-S}^S |S, S_z\rangle \otimes |S, S_z\rangle.  \label{prodex1}
\ee 
Similarly for $\chi=\theta=\phi=\frac{\pi}{2}$, $i.e.$ $(x^1,x^2,x^3, x^4, x^5)=(0,\sinh\rho,0, 0, \cosh\rho)$,  we have 
\be
\hat{H}^{\text{int}}(\rho, \frac{\pi}{2}, \frac{\pi}{2}, \frac{\pi}{2})= -i\frac{\rho}{2} (a^{\dagger}d^{\dagger}-ad) +i\frac{\rho}{2} (b^{\dagger}c^{\dagger}-bc)= \hat{H}^{\text{int}}_{(a,d)}(\rho, 0) +\hat{H}^{\text{int}}_{(b,c)}(\rho, \pi), 
\ee
and 
\begin{align} 
|\text{Sq}\rangle &=|\text{sq} (\rho, 0)\rangle_{a,d} \otimes   |\text{sq} (\rho,\pi)\rangle_{b,c} \nn\\
&=\frac{1}{\cosh^2(\frac{\rho}{2})}\sum_{S=0}^{\infty}\sqrt{2S+1}~(-\tanh(\frac{\rho}{2}))^{2S}\cdot \frac{1}{\sqrt{2S+1}}\sum_{S_z=-S}^S  (-1)^{S-S_z}|S, S_z\rangle \otimes |S, -S_z\rangle.  \label{prodex3}
\end{align}
Each spin sector of (\ref{prodex3}) realizes a maximally entangled spin-singlet state  
\be
\frac{1}{\sqrt{2S+1}}\sum_{S_z=-S}^S (-1)^{S-S_z}|S, S_z\rangle \otimes |S, -S_z\rangle,    \label{maxentspinsing}
\ee
which is utilized in the context of violations of Bell inequality for arbitrary spins  \cite{Garg-Mermin-1982,Peres-1992,Gerry-Albert-2005,Gerry-Benmoussa-HachIII-Albert-2009}.

\subsection{Statistical properties}

\subsubsection{Statistics about spins}

\begin{figure}[tbph]
\center
\includegraphics*[width=140mm]{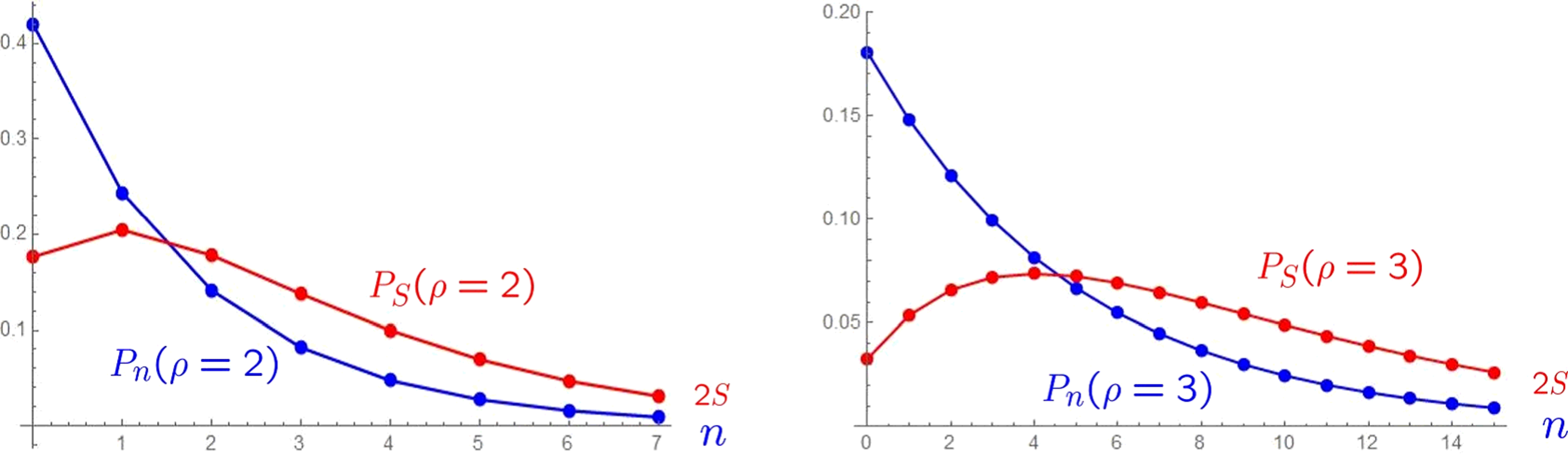}
\caption{Behaviors of the probability functions for the $SO(2,1)$ (blue) and $SO(4,1)$ (red) for  $\rho=2$ and $\rho=3$. The maximum value of $P_S(\rho=2, 3)$ is attained at $2S=1, 4$, while the maximum value of $P_n(\rho=2, 3)$ is attained at $n=0$. For $\rho>\!>1$, the maximum value is attained at $2S \simeq  \frac{1}{4}e^{\rho}$.}
\label{prob.fig}
\end{figure}

The probability function to find  a state  $|S, L_z\rangle \otimes |S, R_z\rangle$ in the $SO(4,1)$ squeezed vacuum is\footnote{
Since product of two D-matrices should be an $SU(2)$ group element, it  satisfies  
\be 
\sum_{L_z, R_z =-S}^S |(D^{(S)}(\phi, \theta, -\chi) ~D^{(S)}(-\chi, -\theta, -\phi))_{L_z, R_z}|^2
 =2S+1.   
\ee 
From the formula, $\sum_{S=0, 1/2, 1, \cdots}  (2S+1) \tanh^{4S} (\frac{\rho}{2}) = \cosh^4(\frac{\rho}{2})$,  one may see  that the summation of (\ref{expprobfunc}) indeed becomes unity: 
\be
\sum_{S=0,1/2,1,3/2,\cdots }\sum_{L_z R_z =-S}^S P_{S, L_z, R_z} (\rho, \chi, \theta, \phi) =1. \label{sumpro1}
\ee 
}   
\begin{align}
P_{S, L_z, R_z} (\rho, \chi, \theta, \phi) &
\equiv |(\langle S, L_z|\otimes \langle S, R_z |)|~\text{Sq}.(\rho, \chi, \theta, \phi)\rangle |^2  \nn\\
&=\frac{1}{\cosh^4(\frac{\rho}{2})}~ \tanh^{4S} (\frac{\rho}{2})
  ~ |(D^{(S)}(\phi, \theta, -\chi) ~D^{(S)}(-\chi, -\theta, -\phi))_{L_z, R_z}|^2.    \label{expprobfunc}
\end{align} 
We also introduce the following probability function to find the two spin-coherent state with $L=R=S$: 
\be
P_S (\rho)\equiv  \sum_{L_z=-S}^S  \sum_{R_z=-S}^S P_{S, L_z, R_z} (\rho, \chi, \theta, \phi) =\frac{2S+1}{\cosh^4(\frac{\rho}{2})}~  \tanh^{4S} (\frac{\rho}{2}).  \label{spintotprob}
\ee 
 Meanwhile, the probability function for the original $SO(2,1)$ squeezed vacuum is (see Appendix \ref{append:basictwomodevac})
\be
P_n(\rho) = \frac{1}{\cosh^2(\frac{\rho}{2})} \tanh^{2n}(\frac{\rho}{2}),   \label{probori}
\ee
which monotonically decreases as the squeeze parameter increases. (\ref{spintotprob}) exhibits non-trivial behavior with respect to $S$, and  
its maximum value  is attained generally at $S\neq 0$ [Fig.\ref{prob.fig}]. This is because of the internal spin degrees of freedom that appears 
as the numerator in  (\ref{spintotprob}).  

It is straightforward to derive 
\begin{align}
&\langle \hat{S}\rangle  = \sum_{S=0,1/2, 1, \cdots} S ~P_S(\rho, \chi, \theta, \phi)  =\sinh^2 (\frac{\rho}{2}), \nn\\ 
&\langle \hat{S}^2\rangle  = \sum_{S=0,1/2, 1, \cdots} S^2 ~P_S(\rho, \chi, \theta, \phi)  =\frac{3}{2}\sinh^4  (\frac{\rho}{2})+\frac{1}{2}\sinh^2  (\frac{\rho}{2}), 
\end{align}
and the variation of spin-magnitude is 
\be
{\langle \Delta \hat{S} ^2\rangle}=\langle \hat{S}^2\rangle -\langle \hat{S}\rangle^2=\frac{1}{8}\sinh^2\rho.  
\ee
The ratio of the spin fluctuation 
\be
\frac{\sqrt{\langle \Delta \hat{S} ^2\rangle}}{\langle \hat{S}\rangle} =\frac{1}{\sqrt{2}}\coth (\frac{\rho}{2}) 
\ee
monotonically decreases from $\infty$ to $\frac{1}{\sqrt{2}}$ as $\rho$ increases from $0$ to $\infty$ (left of Fig.\ref{spins.fig}). 
The expectation values of the left and right spins are  (see Appendix \ref{Sec:so41expval})
\be
\langle \hat{L} \rangle =\langle \hat{R} \rangle 
= \sinh^2(\frac{\rho}{2})=\langle \hat{S} \rangle , 
~~~~\langle \hat{L}^2 \rangle =\langle \hat{R}^2 \rangle 
=\frac{3}{2}\sinh^4 (\frac{\rho}{2}) +\frac{1}{2}\sinh^2(\frac{\rho}{2}) =\langle \hat{S}^2 \rangle . 
\ee
In the expansion of the squeezed vacuum (\ref{spinexpsequso41}), the expansion coefficient in front of the  spin-pair, $\tanh (\frac{\rho}{2})^{2S}$,   will become dominant  for higher $S$ as $\rho$ increases, and so  higher spin states will become significant.    
The correlation between the left and right spins is obtained as 
\be
\langle \hat{L} \cdot \hat{R} \rangle  
=\frac{3}{2}\sinh^4  (\frac{\rho}{2})+\frac{1}{2}\sinh^2  (\frac{\rho}{2})=\langle \hat{S}^2 \rangle .  \label{lrexp}
\ee 
(\ref{lrexp}) can be derived by using the expectation values of the number operators .  
Consequently, $\langle \Delta L^2\rangle= \langle  \hat{L}^2 \rangle -\langle  \hat{L}\rangle^2 $, $\langle \Delta R^2\rangle= \langle  \hat{R}^2 \rangle -\langle  \hat{R}\rangle^2 $ and $\text{cov}(L, R) = \langle  \hat{L}\cdot \hat{R} \rangle -\langle  \hat{L}\rangle  \langle  \hat{R}\rangle$, are given by  
\be 
\langle \Delta L^2\rangle  
=\langle \Delta R^2\rangle 
=\text{cov}(L, R)=\frac{1}{8}\sinh^2\rho ={\langle \Delta \hat{S} ^2\rangle}. 
\ee 
The linear correlation coefficient, $J^{SO(4,1)}(L, R) =\frac{\text{cov}(L, R)}{\sqrt{  \langle \Delta L^2\rangle  ~\langle \Delta R^2\rangle  }}$, takes the maximum value   
\be
J^{SO(4,1)}(L, R) 
=1,  \label{jso41lr}
\ee
which should be compared with that for the $SO(2,1)$ two-mode squeezed vacuum (\ref{numberavevariso21cov}): 
\be
J^{SO(2,1)}(n_a,n_b) =1. \label{jab1}
\ee
As the original  $SO(2,1)$ squeezed vacuum exhibits maximal correlation, the $SO(4,1)$ squeezed vacuum has the maximal correlation between $L$ and $R$ spins. 
\begin{figure}[tbph]
\center
\includegraphics*[width=140mm]{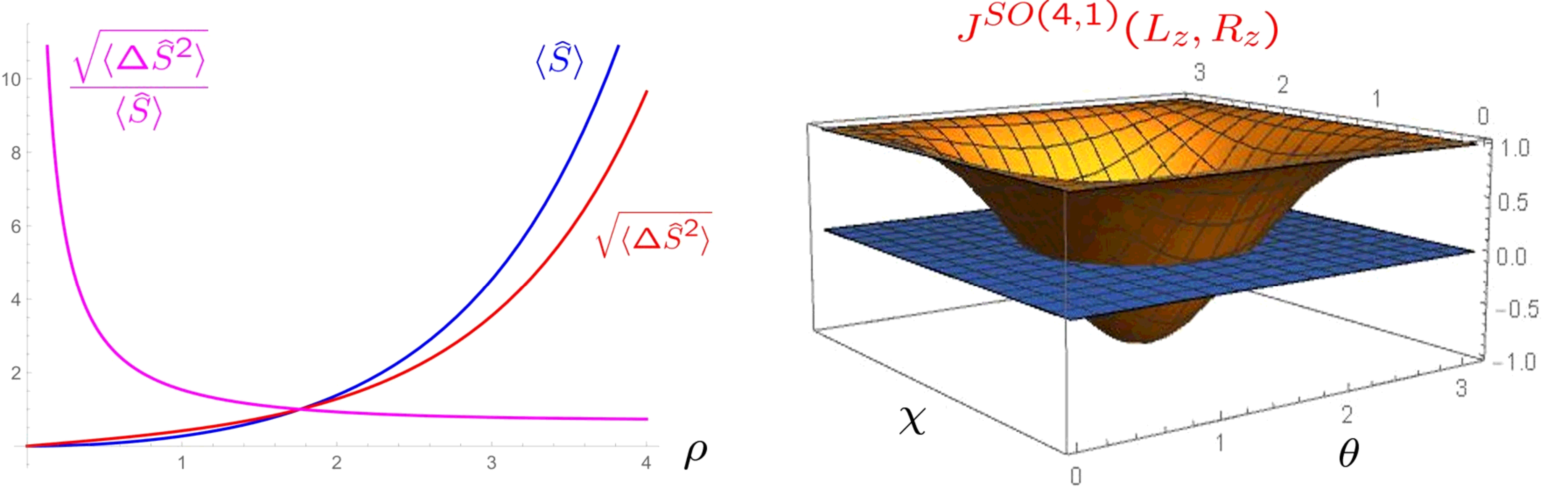}
\caption{Left: Behaviors of the spin magnitude (blue), spin fluctuation (red) and normalized spin fluctuation (magenta) and with respect to the squeeze parameter. 
Right: The angular coordinate dependence of  $J^{SO(4,1)}(L_z, R_z)$.  (The blue plane denotes $J=0$ for comparison.)
}
\label{spins.fig}
\end{figure}
This result is reasonable since, without the spin degrees of freedom, the $SO(4,1)$ squeeze vacuum is almost equivalent to the original $SO(2,1)$ squeezed vacuum  (Fig.\ref{spinexpan.fig}) and two-mode  correlations are analogous in the $SO(4,1)$ and  $SO(2,1)$ states.  For $z$-components of left and right spins,   a bit of calculation shows  '(see Appendix \ref{Sec:so41expval})
\begin{align}
&\langle  {L_z} \rangle = \langle  {R_z} \rangle =0, ~~~~~\langle \Delta {L_z}^2 \rangle = \langle \Delta {R_z}^2 \rangle = \frac{1}{8}\sinh^2 \rho,  \nn\\
&\text{cov}(L_z, R_z) =\langle L_z \cdot R_z\rangle - \langle L_z\rangle   \langle R_z\rangle  = \frac{1}{8}\sinh^2\rho ~(1-2\sin^2\chi\sin^2\theta), 
\end{align}
and then 
\be
J^{SO(4,1)}(L_z, R_z) =\frac{\text{cov}(L_z, R_z)}{\sqrt{  \langle \Delta {L_z}^2\rangle  ~\langle \Delta {R_z}^2\rangle  }}
=1-2\sin^2\chi\sin^2\theta.  \label{jso41lrz} 
\ee
At $\chi=\theta=\frac{\pi}{2}$, the linear correlation (\ref{jso41lrz}) takes maximum negative correlation $J^{SO(4,1)}(L_z, R_z)=-1$, while either at  $\chi=0, \pi$ or $\theta=0, \pi$, (\ref{jso41lrz}) takes the maximum positive correlation $J^{SO(4,1)}(L_z, R_z)=+1$ (Right of Fig.\ref{spins.fig}).  There is  no correlation 
for $\chi$ and $\theta$ that satisfy  $\sin^2\chi~\sin^2\theta=1/2$.

\subsubsection{von Neumann entropy}

\begin{figure}[tbph]
\center
\includegraphics*[width=60mm]{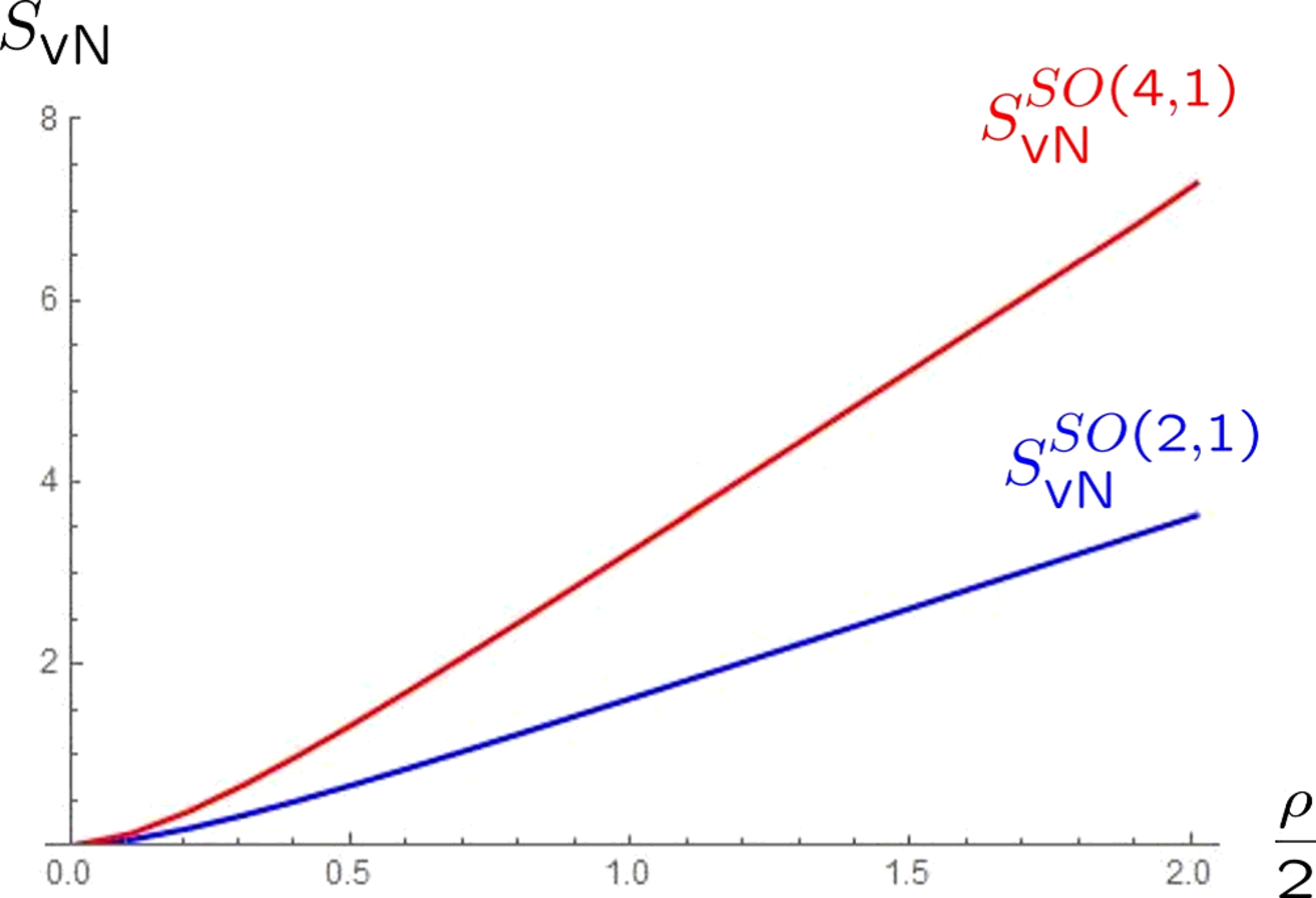}
\caption{The von Neumann entropies of the $SO(2,1)$ (blue) (\ref{oriso21vnentropy}) and $SO(4,1)$ (red) (\ref{so41vnentropy}) squeezed vacua.    }
\label{entropy.fig}
\end{figure}

With the usual definition of  the density operator, 
$\hat{\rho}\equiv |\text{Sq}~(\rho,\chi,\theta,\phi)\rangle \langle \text{Sq}.(\rho,\chi,\theta,\phi) |$,   we construct 
the reduced density operator 
\begin{align}
\hat{\rho}^{(\text{Red})} 
&\equiv 
\sum_{S=0, 1/2, 1, \cdots }\sum_{R_z=-S}^S \langle S, R_z(-\phi,\theta, -\chi)|\hat{\rho}|S, R_z(-\phi, \theta, -\chi)\rangle \nn\\
&=\frac{1}{\cosh^4 (\frac{\rho}{2})} ~\sum_{S=0, 1/2, 1, \cdots} \tanh^{4S}(\frac{\rho}{2}) \sum_{L_z=-S}^S ~ |{S, L_z(\phi, \theta,-\chi)} \rangle \langle {S, L_z(\phi, \theta,-\chi)} |, \label{so41reddens} 
\end{align}
and so the von Neumann entropy is given by 
\be 
S^{SO(4,1)}_{\text{vN}}=-\tr (\hat{\rho}^{(\text{Red})} \log \hat{\rho}^{(\text{Red})}) =\frac{1}{\cosh^4(\frac{\rho}{2})}\sum_{S=0, 1/2, 1. 3/2, \cdots }^{\infty}(2S+1)~ {\tanh^{4S} (\frac{\rho}{2})}~\log \biggl(\frac{\cosh^4(\frac{\rho}{2})}{\tanh^{4S} (\frac{\rho}{2})}\biggr). \label{so41vnentropy}
\ee 
The coefficient $(2S+1)$ on the last right-hand side of (\ref{so41vnentropy}) comes from the $SU(2)$ spin degrees of freedom.  
Meanwhile, the von Neumann entropy of the original squeezed vacuum (\ref{so21vnentropy}) is given by 
\be 
S^{SO(2,1)}_{\text{vN}} 
=\frac{1}{\cosh^2(\frac{\rho}{2})}\sum_{n=0,1,2,\cdots }^{\infty} {\tanh^{2n} (\frac{\rho}{2})}~\log \biggl(\frac{\cosh^2(\frac{\rho}{2})}{\tanh^{2n} (\frac{\rho}{2})}\biggr).  \label{oriso21vnentropy}
\ee 
The von Neumann entropy  of the  $SO(4,1)$ (\ref{so41vnentropy}) grows rapidly than that of the  $SO(2,1)$ (\ref{oriso21vnentropy}), because of the internal spin degrees of freedom (Fig.\ref{entropy.fig}).

\subsection{$SO(4,1)$ uncertainty relations}\label{sec:uncertainty}

From the $SO(4,1)$ Schwinger operators, we  introduce non-commutative 4D coordinates:\footnote{
One can adopt other  4D coordinates:  
\begin{align}
&X_1 =\frac{1}{2\sqrt{2}}(a+a^{\dagger} +b+b^{\dagger}), ~~~X_2 =-i\frac{1}{2\sqrt{2}}(a-a^{\dagger} +b-b^{\dagger}),  \nn\\
&X_3 =\frac{1}{2\sqrt{2}}(c+c^{\dagger} +d+d^{\dagger}), ~~~X_4 =-i\frac{1}{2\sqrt{2}}(c-c^{\dagger} +d-d^{\dagger}),  
\label{anothdeffouxs}
\end{align}
which satisfy same algebra as (\ref{ncrelxs}).  However, 
their expectation values for the $SO(4,1)$ squeezed vacuum are 
\be
\langle X_1 \rangle =  \langle X_2 \rangle  = \langle X_3 \rangle  = \langle X_4 \rangle  = 0,  ~~
\langle (\Delta X_{1})^2 \rangle =  \langle (\Delta X_{2})^2 \rangle  = \langle (\Delta X_{3})^2 \rangle  = \langle (\Delta X_{4})^2 \rangle  = \frac{1}{4}\cosh\rho, 
\ee 
and so 
\be
\langle (\Delta X_{1})^2 \rangle \cdot   \langle (\Delta X_{2})^2 \rangle  =\langle (\Delta X_{3})^2 \rangle \cdot  \langle (\Delta X_{4})^2 \rangle = \frac{1}{16}\cosh^2\rho ~\ge~ \frac{1}{16}. \label{othercoordbound}
\ee
Therefore, for (\ref{anothdeffouxs}), 
only the trivial vacuum ($\rho=0$) saturates the uncertainty bound. 
} 
\begin{align}
&X_1 =\frac{1}{2\sqrt{2}}(a+a^{\dagger} +c+c^{\dagger}), ~~~X_2 =-i\frac{1}{2\sqrt{2}}(a-a^{\dagger} +c-c^{\dagger}),  \nn\\
&X_3 =\frac{1}{2\sqrt{2}}(b+b^{\dagger} +d+d^{\dagger}), ~~~X_4 =-i\frac{1}{2\sqrt{2}}(b-b^{\dagger} +d-d^{\dagger}), 
\end{align}
which satisfy the Heisenberg-Weyl algebra: 
\be
[X_1, X_2] =[X_3, X_4] =i\frac{1}{2}, ~~~[X_1, X_3] =[X_1, X_4]=[X_2, X_3] =[X_2, X_4]=0. \label{ncrelxs}
\ee 
From the expectation values for the squeezed vacuum (see Appendix \ref{appendix:expectationval}), we have 
\be
\langle X_1 \rangle =  \langle X_2 \rangle  = \langle X_3 \rangle  = \langle X_4 \rangle  = 0, 
\ee
and 
the variants  
\be
\langle (\Delta X_{m})^2 \rangle \equiv \langle  (X_{m})^2 \rangle-{\langle  X_{m} \rangle}^2=\langle  (X_{m})^2 \rangle~~~~(m=1,2,3,4)
\ee
as 
\begin{align}
&\langle (\Delta X_1)^2 \rangle=\langle (\Delta X_3)^2 \rangle   =\frac{1}{4}(\cosh \rho-\sinh\rho\cos\chi), \nn\\
&\langle (\Delta X_2)^2 \rangle =\langle (\Delta X_4)^2 \rangle   =\frac{1}{4}(\cosh \rho+\sinh\rho\cos\chi).  \label{fourmodevariants}
\end{align}
Consequently, 
\be 
\langle (\Delta X_1)^2 \rangle \langle (\Delta X_2)^2 \rangle =  \langle (\Delta X_3)^2 \rangle \langle (\Delta X_4)^2 \rangle =\frac{1}{16}+\frac{1}{16}\sinh^2\rho~\sin^2\chi ~\ge ~\frac{1}{16} .  \label{uncertaintyrel}
\ee
As the squeeze parameter $\rho$ increases, the uncertainty (\ref{uncertaintyrel}) monotonically increases.  
The inequality is saturated at $\chi=0, \pi$: 
\be
\langle (\Delta X_1)^2 \rangle  =\frac{1}{4}e^{\mp \rho}, ~~~\langle (\Delta X_2)^2 \rangle  =\frac{1}{4}e^{\pm \rho}, ~~\langle (\Delta X_3)^2 \rangle  =\frac{1}{4}e^{\mp \rho}, ~~~\langle (\Delta X_4)^2 \rangle  =\frac{1}{4}e^{\pm \rho}.
\ee

\begin{figure}[tbph]
\center
\includegraphics*[width=130mm]{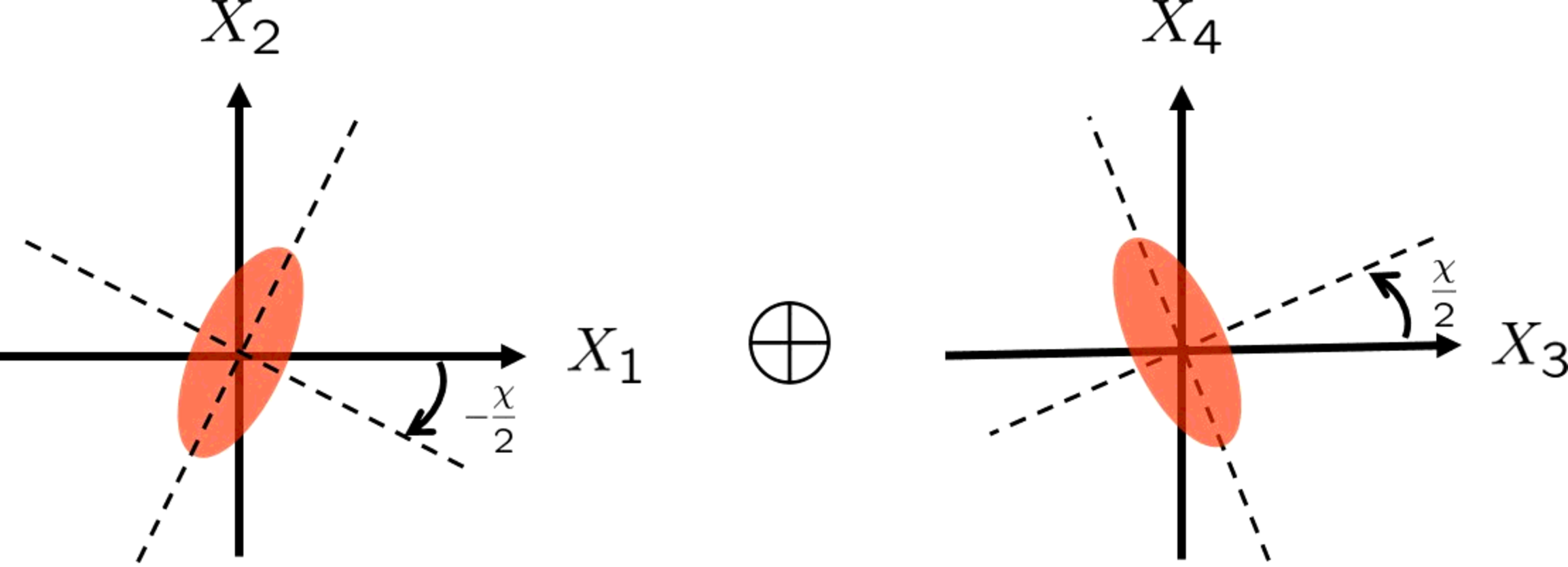}
\caption{The uncertainty regions for $\theta=0$.  }
\label{4duncertain.fig}
\end{figure}

To remove the $\chi$-dependence in (\ref{fourmodevariants}),  let us apply the following coordinate rotation (same as in the original case (\ref{rotatredef})): 
\be
\begin{pmatrix}
X_1 \\
X_2 
\end{pmatrix} ~~\rightarrow~~
\begin{pmatrix}
\cos\frac{\chi}{2} & -\sin\frac{\chi}{2} \\
\sin\frac{\chi}{2}  & \cos\frac{\chi}{2}
\end{pmatrix} \begin{pmatrix}
X_1 \\
X_2 
\end{pmatrix},~~~\begin{pmatrix}
X_3 \\
X_4 
\end{pmatrix} ~~\rightarrow~~
\begin{pmatrix}
\cos\frac{\chi}{2} & \sin\frac{\chi}{2} \\
-\sin\frac{\chi}{2}  & \cos\frac{\chi}{2}
\end{pmatrix} \begin{pmatrix}
X_3 \\
X_4 
\end{pmatrix}, 
\ee
but we end up with
\begin{align}
&\langle (\Delta X_1)^2 \rangle  =\langle (\Delta X_3)^2 \rangle  =\frac{1}{4}(\cosh \rho -\sinh\rho (\cos^2\chi+\sin^2\chi\cos\theta)), \nn\\
&\langle (\Delta X_2)^2 \rangle  =\langle (\Delta X_4)^2 \rangle  =\frac{1}{4}(\cosh \rho +\sinh\rho (\cos^2\chi+\sin^2\chi\cos\theta)).   \label{resultsrotationnoncom}
\end{align}
As mentioned  around (\ref{sumanglesphi}),   we can remove $\phi$-dependence  by the redefinition of the Schwinger operators but not $\chi$, so the angle dependence in (\ref{fourmodevariants}) is not generally removal unlike the original $SO(2,1)$ case (see Appendix \ref{append:basictwomodevac}).   However, 
for the special case  $\theta=0$, $\chi$-dependence in (\ref{resultsrotationnoncom}) can be removed: 
\be
\langle (\Delta X_1)^2 \rangle  =\langle (\Delta X_3)^2 \rangle~\rightarrow~ \frac{1}{4}e^{-\rho},~~~\langle (\Delta X_2)^2 \rangle  =\langle (\Delta X_4)^2 \rangle~\rightarrow~ \frac{1}{4}e^{\rho}.
\ee
Recall that for $\theta=0$ the $SO(4,1)$ is simply reduced to the direct product of two $SO(2,1)$  squeezed states with  opposite angles, $\chi$ and $-\chi$ (\ref{tbsqueezedstates}), and in each $SO(2,1)$ squeezed vacuum, the minimum uncertainty is realized (Fig.\ref{4duncertain.fig}). 

\section{Summary}\label{sec:summary}

\begin{table}
\center 
   \begin{tabular}{|c|c|c|}\hline
        &  $SO(2,1)$ squeezed vacuum  &   $SO(4,1)$ squeezed vacuum   \\ \hline
Symmetry group        &  $SO(2,1) = Sp(2; \mathbb{R}) =U(1; \mathbb{H}')$       &  $SO(4,1) =U(1,1; \mathbb{H})$      \\ \hline   
 Representation    &  Majorana/Dirac   &  Dirac        \\ \hline
   Topological map    &  1st non-compact Hopf map   &  2nd non-compact (hybrid) Hopf map        \\ \hline
    Quantum manifold      &  Bloch two-hyperboloid  $H^{2}$   &  Bloch four-hyperboloid  $H^4$   \\ \hline 
     Degrees of freedom    &   One/two-mode  photons   &  Four-mode photons or two-mode spins     \\ \hline 
    \end{tabular}       
\caption{ Comparison between the $SO(2,1)$ squeezed vacuum and the $SO(4,1)$ squeezed vacuum.    
}
\label{table:correspDNYM}
\end{table}

In this work, we proposed a generalized framework of the squeezed state that includes the spin degrees of freedom 
based on the  Bloch four-hyperboloid (Table \ref{table:correspDNYM}).  
 While the obtained $SO(4,1)$ squeezed vacuum is a four-mode squeezed state in the photon picture, it can  be  interpreted as an entangled spin state with the  Schwinger's angular momentum operator formalism. 
The four parameters of the $SO(4,1)$ squeezed vacuum have a clear geometric origin in  the Bloch four-hyperboloid: 
The squeeze parameter   corresponds to the radial coordinate  and the three internal spin parameters correspond to  
the three angular coordinates on $S^3$-latitude. 
As the original $SO(2,1)$ squeezed vacuum is expressed by a superposition of identical number two-mode photons, 
the $SO(4,1)$ squeezed vacuum is realized as a superposition of  identical  magnitude spin-pairs.  
 The $SO(4,1)$ squeezed vacuum exhibits a hybrid entanglement:    
One  is about the infinite set of the spin-pairs and  the other is  about  two spins in each spin-pair. 
The linear correlation coefficient between the left and right spins realizes the maximum correlation. 
The statistical properties of the $SO(4,1)$ squeezed vacuum  are qualitatively different to  the original $SO(2,1)$ squeezed vacuum in several respects, such as
 the  spin magnitude dependence of  the probability function and the angular coordinate dependence of  the linear correlation between the left and right third-spin-components.     The uncertainty relation was also generalized in 4D with coordinate dependence of the Bloch four-hyperboloid.      

     The Bloch four-hyperboloid has provided a  formalism that accommodates higher spins    
and  squeezed states on the same footing.   
 Since  non-compact geometry can incorporate compact geometry  as its internal structure,  theory based on non-compact geometry generally provides  a more comprehensive framework than that  on  compact geometry.  
While the importance of  compact geometry in quantum information has already been appreciated  \cite{Bengtsson-Zyczkowski-2006},  non-compact or indefinite signature geometry has  less been exploited so far.  
We hope that  significance of   non-compact  geometry  in    quantum information will be further unveiled in  future developments.

\section*{Acknowledgements}

 This work was supported by JSPS KAKENHI Grant Number~16K05334 and 16K05138.

\appendix

\section{Basic properties of the $SO(2,1)$ two-mode squeezed vacuum }\label{append:basictwomodevac}

This section is mainly based on Sec.7.7 of Gerry and Knight \cite{Gerry-Knight-2004}.

\subsection{One-mode $SO(2,1)$ squeezed vacuum (Majorana representation)}

For the one-mode  $SO(2,1)$ squeezed vacuum 
\be 
|\text{sq}(\rho, \phi)\rangle = e^{-\frac{\rho}{4}e^{i\phi}{a^{\dagger}}^2 +\frac{\rho}{4}e^{-i\phi}{a}^2 }|0\rangle,
\ee
 we have 
\be
\langle \hat{n}\rangle = \sinh^2(\frac{\rho}{2}), ~~~~\langle \hat{n}^2\rangle = \sinh^4(\frac{\rho}{2}) + \frac{1}{2}\sinh^2{\rho}.  \label{ondemoparticlenum}
\ee
The variant of the number is 
\be
\langle {\Delta \hat{n}}^2\rangle =\frac{1}{2}\sinh^2\rho,   \label{ondemodevariant}
\ee
and then 
\be
\frac{\sqrt{\langle {\Delta n}^2\rangle}}{\langle n_a\rangle} =\sqrt{2}\coth(\frac{\rho}{2})~\ge \sqrt{2}.
\ee
The non-commutative coordinates 
which satisfy 
\be
[X_1, X_2]=i\frac{1}{2} 
\ee
are defined as 
\be
X_1=\frac{1}{2}(a+a^{\dagger}), ~~X_2=-i\frac{1}{2}(a-a^{\dagger}). 
\ee
The relevant expectation values are 
\be
\langle X_1 \rangle = \langle X_2 \rangle =0,  ~~~\langle (\Delta X_1)^2 \rangle = \frac{1}{4}(\cosh \rho-\sinh\rho\cos\phi), ~~\langle (\Delta X_2)^2 \rangle = \frac{1}{4}(\cosh \rho+\sinh\rho\cos\phi). \label{1modex1x2squasq}
\ee 
An appropriate choice of the rotation 
\be
\begin{pmatrix}
X_1 \\
X_2
\end{pmatrix} ~~\rightarrow~~
\begin{pmatrix}
\cos\frac{\phi}{2} & -\sin\frac{\phi}{2} \\
 \sin\frac{\phi}{2} & \cos\frac{\phi}{2} 
\end{pmatrix}  \begin{pmatrix}
X_1 \\
X_2
\end{pmatrix}, \label{rotatredef}
\ee
can remove the angle dependence of the variants: 
\be
\langle (\Delta X_1)^2 \rangle = \frac{1}{4}e^{-\rho}, ~~\langle (\Delta X_2)^2 \rangle = \frac{1}{4}e^{\rho}. \label{rotresuori}
\ee 

\subsection{Two-mode $SO(2,1)$ squeezed vacuum (Dirac representation)}

For the $SO(2,1)$ two-mode squeezed vacuum 
\be
|\text{sq}(\rho, \phi)\rangle = e^{-\frac{\rho}{2}e^{i\phi}{a^{\dagger}}b^{\dagger} +\frac{\rho}{2}e^{-i\phi}{a}b }|0\rangle,
\ee
the probability function is given by  
\be
P_{n_a, n_b}(\rho) =|\langle n_a, n_b| \text{sq}(\rho, \phi) \rangle |^2 =\delta_{n_a, n_b} \cdot \frac{1}{\cosh^2 (\frac{\rho}{2})} \tanh^{2n_a} (\frac{\rho}{2}), 
\ee
which satisfies 
\be
\sum_{n_a, n_b=0}^{\infty}~P_{n_a, n_b}(\rho) = \sum_{n=0}^{\infty}P_{n}(\rho)=1.  
\ee
Here, $P_n$ denotes the probability to find $|n\rangle \otimes |n\rangle$ in the squeezed vacuum: 
\be
P_{n}(\rho) \equiv P_{n, n}(\rho)  =\frac{1}{\cosh^2 (\frac{\rho}{2})}~ \tanh^{2n}(\frac{\rho}{2}).  \label{so21probfun}
\ee
The number averages are evaluated as 
\begin{subequations}
\begin{align}
&\langle \hat{n}_a\rangle = \langle \hat{n}_b\rangle = \sum_{n=0}^{\infty} n \cdot P_{n} (\rho, \phi) =\sinh^2(\frac{\rho}{2}),  
 \\
&\langle {\hat{n}_a}^2\rangle = \langle {\hat{n}_b}^2\rangle = \langle {\hat{n}_a} \hat{n}_b \rangle = \sum_{n=0}^{\infty} n^2 \cdot P_{n} (\rho, \phi) =\sinh^4(\frac{\rho}{2}) + \frac{1}{4}\sinh^2{\rho},  
\end{align} \label{numberaveso21} 
\end{subequations}
and the variants and covariants are  
\be
\langle {\Delta n_a}^2\rangle = \langle {\Delta n_b}^2\rangle = \text{cov}(n_a, n_b)  =\frac{1}{4}\sinh^2 \rho. \label{nanbfluc}
\ee 
The ratio of the number fluctuation and the  linear correlation are respectively given by 
\begin{subequations}
\begin{align}
&\frac{\sqrt{\langle {\Delta n_a}^2\rangle}}{\langle n_a\rangle} =\coth(\frac{\rho}{2})~\ge 1, \label{rationns}\\
&J(n_a, n_b) =\frac{ \text{cov}(n_a, n_b)  }{\sqrt{\langle {\Delta n_a}^2\rangle ~ \langle {\Delta n_b}^2\rangle}}=1. \label{numberavevariso21cov}
\end{align}
\end{subequations}
The ratio (\ref{rationns}) monotonically decreases to unity, as the squeeze parameter increases. 
The statistics of the number  is  a super-Poissonian statistics. 

With the density operator $\hat{\rho}\equiv |\text{sq}(\rho, \phi)\rangle \langle \text{sq}(\rho, \phi)|$,  
the reduced density operator is defined as
\be
\hat{\rho}^{(\text{Red})} 
\equiv \text{tr}_R \hat{\rho} 
=\frac{1}{\cosh^2 (\frac{\rho}{2})} ~\sum_{n=0,  1, 2,  \cdots} \tanh^{2n}(\frac{\rho}{2})  |n \rangle_L \langle n |_L, 
\ee 
and the associated von Neumann entropy is  
\be 
S_{\text{vN}}=-\tr (\hat{\rho}^{(\text{Red})} \log \hat{\rho}^{(\text{Red})}) =\frac{1}{\cosh^2(\frac{\rho}{2})}\sum_{n=0}^{\infty} {\tanh^{2n} (\frac{\rho}{2})}~\log \biggl(\frac{\cosh^2(\frac{\rho}{2})}{\tanh^{2n} (\frac{\rho}{2})}\biggr). \label{so21vnentropy}
\ee 
Expectation values for the non-commutative coordinates 
\be
X_1 =\frac{1}{2\sqrt{2}} (a+a^{\dagger}+b+b^{\dagger}), ~~~X_2 =-i\frac{1}{2\sqrt{2}}(a-a^{\dagger} +b-b^{\dagger}),  
\ee
are given by 
\be
\langle X_1 \rangle = \langle X_2 \rangle =0,  ~~~\langle (\Delta X_1)^2 \rangle = \frac{1}{4}(\cosh \rho-\sinh\rho\cos\phi), ~~\langle (\Delta X_2)^2 \rangle = \frac{1}{4}(\cosh \rho+\sinh\rho\cos\phi),  \label{x1x2squasq}
\ee 
which are identical to the one-mode case (\ref{1modex1x2squasq}).

\section{Expectation values of the $SO(4,1)$ and $SO(2,3)$ squeezed vacua}\label{appendix:expectationval}

Even without knowing a simple number state expansion of squeezed vacuum, 
we can readily evaluate squeezed vacuum expectation values by the covariance of the Schwinger operator: 
\be
\hat{S}^{\dagger}~\hat{\Psi}~ \hat{S} =M\hat{\Psi},   
\ee
which implies  the following useful formula  
\be
\langle O(\hat{\Psi}) \rangle=\langle \text{Sq} |O(\hat{\Psi}) |\text{Sq}\rangle =\langle 0|\hat{S}^{\dagger} O(\hat{\Psi}) \hat{S}|0\rangle=\langle 0|O( \hat{S}^{\dagger}\hat{\Psi} \hat{S}) |0\rangle =\langle 0|O( M\hat{\Psi}) |0\rangle.  
\ee

\subsection{$SO(4,1)$ squeezed vacuum}\label{Sec:so41expval} 

For the $SO(4,1)$ squeezed vacuum (\ref{simpledefsquso41}),    
we can  derive 
\begin{align}
&\langle 0| \hat{S}^{\dagger}a^2 \hat{S}|0\rangle = \langle 0| \hat{S}^{\dagger}c^2 \hat{S} |0\rangle =\langle 0| \hat{S}^{\dagger}a c^{\dagger} \hat{S} |0\rangle =0, \nn\\
&\langle 0| \hat{S}^{\dagger}aa^{\dagger} \hat{S} |0\rangle =|M_{11}|^2+|M_{12}|^2 =\cosh^2(\frac{\rho}{2}), ~~~ \langle 0| \hat{S}^{\dagger}cc^{\dagger} \hat{S} |0\rangle =|M_{33}|^2+|M_{34}|^2 =\cosh^2(\frac{\rho}{2}), \nn\\
&\langle 0| \hat{S}^{\dagger}a^{\dagger} a \hat{S} |0\rangle =|M_{13}|^2+|M_{14}|^2 =\sinh^2(\frac{\rho}{2}), ~~~ \langle 0| \hat{S}^{\dagger}c^{\dagger}c \hat{S} |0\rangle =|M_{31}|^2+|M_{32}|^2 =\sinh^2(\frac{\rho}{2}), \nn\\
&\langle 0| \hat{S}^{\dagger} ac \hat{S} |0\rangle =M_{11} M_{31}^*+M_{12}M_{32}^* =-\frac{1}{2}\sinh\rho (\cos\chi+i\sin\chi\cos\theta),   
\end{align}
where  
$M(\rho, \chi,\theta,\phi)$ is  (\ref{squeezematexp}). 
With these results, the expectation value of  
\be
(X_1)^2 =\frac{1}{8}(a+a^{\dagger}+c+c^{\dagger})^2 =\frac{1}{8}(a^2+aa^{\dagger}+a^{\dagger}a +{a^{\dagger}}^2+c^2+cc^{\dagger}+c^{\dagger}c +{c^{\dagger}}^2 + 2ac +2ac^{\dagger}+2a^{\dagger}c+2a^{\dagger}c^{\dagger}), 
\ee
is evaluated as 
\be
\langle (X_1)^2 \rangle   = \frac{1}{4}(\cosh \rho-\sinh\rho\cos\chi). 
\ee
In a similar manner, we have 
\begin{align}
&\langle \hat{n}_a \rangle =\langle \hat{n}_b \rangle =\langle \hat{n}_c \rangle =\langle \hat{n}_d \rangle =\sinh^2(\frac{\rho}{2}),\nn \\
&\langle {\hat{n}_a}^2 \rangle =\langle {\hat{n}_b}^2 \rangle =\langle {\hat{n}_c}^2 \rangle =\langle {\hat{n}_d}^2 \rangle =2\sinh^4(\frac{\rho}{2}) +\sinh^2(\frac{\rho}{2}), \nn\\
&\langle {\Delta n_a}^2 \rangle =\langle {\Delta n_b}^2 \rangle =\langle {\Delta n_c}^2 \rangle =\langle {\Delta n_d}^2 \rangle =\frac{1}{4}\sinh^2{\rho}, 
\end{align} \label{numberavevariso41}
which  are exactly equal to (\ref{numberaveso21}) and (\ref{nanbfluc}). 
Furthermore with 
\begin{align}
&\langle \hat{n}_a \hat{n}_b\rangle =\langle \hat{n}_c \hat{n}_d\rangle  =\sinh^4(\frac{\rho}{2}), \nn\\
&\langle \hat{n}_a \hat{n}_c\rangle  =\langle \hat{n}_b \hat{n}_d\rangle  =\frac{1}{4}\sinh^2\rho ~(\cos^2\chi +\sin^2\chi\cos^2\theta)+\sinh^4(\frac{\rho}{2}), \nn\\
&\langle \hat{n}_a \hat{n}_d\rangle  =\langle \hat{n}_b \hat{n}_c\rangle=\frac{1}{4}\sinh^2\rho ~\sin^2\chi\sin^2\theta+\sinh^4(\frac{\rho}{2}).  
\end{align}
the covariances,   
$\text{cov} (n_a, n_b) =\langle \hat{n}_a \hat{n}_b\rangle -\langle \hat{n}_a\rangle \langle \hat{n}_b\rangle$,   
are  derived as 
\begin{align}
&\text{cov}( n_a, n_b) =\text{cov}( n_c, n_d)  =0, \nn\\
&\text{cov}(  n_a, n_c) =\text{cov}(  n_b, n_d)  =\frac{1}{4}\sinh^2\rho ~(1-\sin^2\chi\sin^2\theta), \nn\\
&\text{cov}(  n_a, n_d) =\text{cov}(  n_b, n_c) =\frac{1}{4}\sinh^2\rho ~\sin^2\chi\sin^2\theta,
\end{align}
The linear correlation coefficients,  
$J(n_a, n_b) =\frac{1}{\sqrt{\langle {\Delta n_a}^2 \rangle \langle {\Delta n_b}^2 \rangle }}~\text{cov}( n_a,  n_b)$,  
are 
\begin{align}
&J( n_a, n_b) =J( n_c, n_d)  =0, \nn\\
&J(  n_a, n_c)  =J(  n_b, n_d)  =1-\sin^2\chi\sin^2\theta, \nn\\
&J(  n_a, n_d) =J(  n_b, n_c)= \sin^2\chi\sin^2\theta. \label{numberavevariso41cov}
\end{align} 
The linear correlation coefficients  (\ref{numberavevariso41cov}) depend on the angular coordinates, while for the original squeezed vacuum they  are constant (\ref{numberavevariso21cov}). The linear correlation coefficients  (\ref{numberavevariso41cov})  become the maximum value 1 at $\chi=0, \pi$ or $\theta=0, \pi$, while they vanish either at $\chi=\theta=\frac{\pi}{2}$.

The expectation values for the left and right spins are  evaluated as 
\be
\langle \hat{L} \rangle =\langle \hat{R} \rangle = \frac{1}{2}\langle \hat{n}_a\rangle + \frac{1}{2}\langle \hat{n}_b\rangle
 = \sinh^2(\frac{\rho}{2}), ~~~~\langle \hat{L}_z \rangle =\langle \hat{R}_z \rangle =  0, 
\ee
and 
\begin{align}
&\langle \hat{L}^2 \rangle =\langle \hat{R}^2 \rangle =\frac{1}{4}\langle (\hat{n}_a+\hat{n}_b)^2\rangle =\frac{3}{2}\sinh^4 (\frac{\rho}{2}) +\frac{1}{2}\sinh^2(\frac{\rho}{2}), \nn \\
&\langle \hat{L} \cdot \hat{R} \rangle =  \frac{1}{4}\langle (\hat{n}_a+\hat{n}_b) (\hat{n}_c+\hat{n}_d)\rangle =\frac{1}{8}\sinh^2\rho +\sinh^4(\frac{\rho}{2}), \nn\\
&\langle \hat{L}_z^2 \rangle =\langle \hat{R}_z^2 \rangle =\frac{1}{4}\langle (\hat{n}_a-\hat{n}_b)^2\rangle =\frac{1}{8}\sinh^2 {\rho}, \nn \\
&\langle \hat{L}_z \cdot \hat{R}_z \rangle =  \frac{1}{4}\langle (\hat{n}_a-\hat{n}_b) (\hat{n}_c-\hat{n}_d)\rangle =\frac{1}{8}\sinh^2\rho~(1 -2\sin^2\chi~ \sin^2\theta). 
\end{align}
Consequently, 
\begin{align}
\langle \Delta L^2\rangle  &  =\langle  \hat{L}^2 \rangle -\langle  \hat{L}\rangle^2 =\frac{1}{4}(\langle \Delta n_a^2\rangle +2\text{cov}(n_a,n_b)+\langle \Delta n_b^2\rangle  ) =\frac{1}{8}\sinh^2\rho =\langle \Delta R^2\rangle, \nn\\ 
\langle \Delta L_z^2\rangle  &=\langle  \hat{L}_z^2 \rangle -\langle  \hat{L}_z\rangle^2 
=\frac{1}{8}\sinh^2\rho =\langle \Delta R_z^2\rangle, 
\end{align}
and 
\begin{align}
\text{cov}(L, R) &= \langle  \hat{L}\cdot \hat{R} \rangle -\langle  \hat{L}\rangle  \langle  \hat{R}\rangle=\frac{1}{4}(\text{cov}(n_a, n_c) +\text{cov}(n_a, n_d) +\text{cov}(n_b, n_c) +\text{cov}(n_b, n_d)  )  \nn\\
&=\frac{1}{8}\sinh^2\rho , \nn\\    
\text{cov}(L_z, R_z) &= \langle  \hat{L}_z\cdot \hat{R}_z \rangle -\langle  \hat{L}_z\rangle  \langle  \hat{R}_z\rangle=\frac{1}{4}(\text{cov}(n_a, n_c) -\text{cov}(n_a, n_d) -\text{cov}(n_b, n_c) +\text{cov}(n_b, n_d)  )  \nn\\
&=\frac{1}{8}\sinh^2\rho ~(1-2\sin^2\chi~\sin^2\theta).
\end{align}

\subsection{$SO(2,3)\simeq Sp(4; \mathbb{R})$ squeezed vacua}

For comparison,  
 we also derive expectation values for the $SO(2,3)$ squeezed vacua \cite{Hasebe-2019}.   
In the $SO(2,3)$ case, we have two-mode  and four-mode squeezed states, since $SO(2,3)$ allows Majorana representation and  Dirac representation.  
The $SO(2,3)$ squeezed vacua are also physically distinct for the Dirac-type and the Schwinger-type.  
In this section, we utilize the following parameterization of the coordinates on $H^{2,2}$: 
\begin{align}
&x^1=\sin\theta  \cos\chi\sinh \rho, ~~x^2=\sin\theta \sin\chi\sinh \rho, ~~x^3= \cos\phi \sin\theta \cosh \rho, \nn\\
&x^4= \sin\phi\sin\theta\cosh\rho ,~~~x^5=\cos\theta,  
\end{align}
which satisfy 
\be
(x^1)^2+ (x^2)^2- (x^3)^2- (x^4)^2- (x^5)^2 =-1.
\ee

\subsubsection{Two-mode $SO(2,3)$ squeezed vacuum (Majorana representation)}

\begin{itemize}
\item{Dirac-type}
\end{itemize}

\begin{align}
&\langle \hat{n}_a\rangle =\langle \hat{n}_b\rangle =\sinh^2{\rho} ~\sin^2(\frac{\theta}{2}), \nn\\
&\langle {\Delta n_a}^2\rangle = \langle {\Delta n_b}^2\rangle =  \sinh^2\rho ~\sin^2(\frac{\theta}{2})\biggl(1+\cosh(2\rho)\sin^2(\frac{\theta}{2})\biggr), \nn\\
&\text{cov}(n_a, n_b) =
\frac{1}{4}\sinh^2\rho~\sin^2\theta , \nn \\
&J(n_a, n_b) =\frac{1}{4(1+\cosh(2\rho)~\sin^2(\frac{\theta}{2}))}.  \label{diracso232mode}
\end{align}
Similar to  the  $SO(4,1)$ case (see Appendix \ref{Sec:so41expval}), the expectation values for the Dirac-type $SO(2,3)$ squeezed vacuum have angle dependence but in a  different manner. 

\begin{itemize}
\item{Schwinger-type}
\end{itemize}

\begin{align}
&\langle \hat{n}_a\rangle =\langle \hat{n}_b\rangle =\sinh^2(\frac{\rho}{2}), \nn\\
&\langle {\Delta n_a}^2\rangle = \langle {\Delta n_b}^2\rangle =  \langle {\Delta n_b}^2\rangle =  \frac{1}{2}\sinh^2\rho, \nn\\
&\text{cov}(n_a, n_b) =0 , \nn \\
&J(n_a, n_b) =0.  \label{schwingerso232mode}
\end{align}
The Schwinger-type $SO(2,3)$ squeezed vacuum is just given by  a direct product of two $SO(2,1)$ squeezed vacuum states \cite{Hasebe-2019}.  Therefore,  (\ref{schwingerso232mode}) 
is just given by two copies  of (\ref{ondemoparticlenum}) and (\ref{ondemodevariant}), and there is no correlation between $a$ and $b$ modes.

\subsubsection{Four-mode $SO(2,3)$ squeezed vacuum (Dirac representation) }

\begin{itemize}
\item{Dirac-type}
\end{itemize}

\begin{align}
&\langle \hat{n}_a\rangle =\langle \hat{n}_b\rangle=\langle \hat{n}_c\rangle =\langle \hat{n}_d\rangle =\sinh^2{\rho}~\sin^2(\frac{\theta}{2}), \nn\\
&\langle {\Delta n_a}^2\rangle = \langle {\Delta n_b}^2\rangle =\langle {\Delta n_c}^2\rangle = \langle {\Delta n_d}^2\rangle =  \sinh^2\rho~\sin^2(\frac{\theta}{2})(1+\sin^2\rho~\sin^2(\frac{\theta}{2})), \nn\\
&\text{cov}(n_a, n_b) =\text{cov}(n_c, n_d) = \frac{1}{4}\sinh^2(2\rho)\sin^4(\frac{\theta}{2}), ~~~\text{cov}(n_a, n_c) =\text{cov}(n_b, n_d)=0,~~~ \nn\\
&\text{cov}(n_a, n_d) =\text{cov}(n_b, n_c) =\frac{1}{4}\sinh^2\rho~\sin^2\theta, \nn\\
&J(n_a, n_b) =J(n_c, n_d) = \frac{1}{1+\sinh^2\rho~\sin^2(\frac{\theta}{2})}\cosh^2\rho~\sin^2(\frac{\theta}{2}), ~~~J(n_a, n_c) =J(n_b, n_d)=0,\nn\\
&J(n_a, n_d) =J(n_b, n_c) =\frac{1}{1+\sinh^2\rho~\sin^2(\frac{\theta}{2})}~\cos^2(\frac{\theta}{2}).  \label{diracso234mode}
\end{align}
As in the two-mode case (\ref{diracso232mode}), (\ref{diracso234mode}) exhibits non-trivial angle dependence but in a different manner. As $\rho$ increases, $J(n_a, n_b)$ and $J(n_c, n_d)$  monotonically increase while $J(n_a, n_d)$ and $J(n_b, n_c)$ monotonically decrease. 

\begin{itemize}
\item{Schwinger-type}
\end{itemize}

\begin{align}
&\langle \hat{n}_a\rangle =\langle \hat{n}_b\rangle=\langle \hat{n}_c\rangle =\langle \hat{n}_d\rangle =\sinh^2(\frac{\rho}{2}),   \nn\\
&\langle {\Delta n_a}^2\rangle = \langle {\Delta n_b}^2\rangle =\langle {\Delta n_c}^2\rangle = \langle {\Delta n_d}^2\rangle =  \frac{1}{4}\sinh^2\rho, \nn\\
&\text{cov}(n_a, n_b) =\text{cov}(n_c, n_d) = 1, ~~~\text{cov}(n_a, n_c) =\text{cov}(n_a, n_d) =\text{cov}(n_b, n_c) =\text{cov}(n_b, n_d) =0, \nn\\
&J(n_a, n_b) =J(n_c, n_d) = 1, ~~~J(n_a, n_c) =J(n_a, n_b) =J(n_b, n_c) =J(n_b, n_d) =0. 
\end{align}
This is a straightforward generalization of the  two-mode case (\ref{schwingerso232mode}). 




\end{document}